\documentclass[useAMS,usenatbib]{mn2e}
\usepackage{graphicx} 
\usepackage{times}
\usepackage{epsfig}
\usepackage{rotating}
\usepackage{sidecap}

\def\lesssim{\mathrel{\hbox{\rlap{\hbox{\lower4pt\hbox{$\sim$}}}\hbox{$<$}}}}
\let\la=\lesssim
\def\gtrsim{\mathrel{\hbox{\rlap{\hbox{\lower4pt\hbox{$\sim$}}}\hbox{$>$}}}}
\let\ga=\gtrsim

\title[Outburst evolution of  H1743-322]{On the outburst evolution of H1743-322: a 2008/2009 comparison}
\author[Motta et al.]{S.~Motta$^{1,2}$, T.~Mu\~noz-Darias$^{1}$, T.~Belloni$^{1}$ \\
$^{1}$INAF-Osservatorio Astronomico di Brera, Via E. Bianchi 46, I-23807 Merate (LC), Italy\\
$^{2}$Universit\`a dell'Insubria, Via Valleggio 11, I-22100 Como, Italy \\
}
\begin{document}
\maketitle
\begin{abstract}
We present two observational campaigns  performed with the RXTE satellite on the black hole transient H 1743-322. The source was observed in outburst on two separate occasions between October-November 2008 and May-July 2009. We have carried out timing and spectral analysis of the data set, obtaining a complete state classification of all the observations. We find that all the observations are well described by using a spectral model consisting of a disk-blackbody, a powerlaw + reflection + absorption and a gaussian emission component. During the 2009 outburst the system followed the canonical evolution through all the states seen in black hole transients. In the 2008 outburst only the hard states were reached. 

The early evolution of the spectral parameters is consistent between the two epochs, and it does not provide clues about the subsequent behavior of the source. 
%Only the high-energy cutoff seems higher during the 2008 outburst, suggesting a higher coronal temperature that could affect the accretion dynamics. 
The variation of the flux associated to the two main spectral components (i.e. disk and powerlaw) allows us to set a lower limit to the orbital inclination of the system of $\geq$ 43$^{\circ}$.
\end{abstract}

\section{Introduction}
Black holes X-ray transients (BHTs) spend most of their lives in \textquotedblleft quiescence\textquotedblright, displaying very low luminosities ($\sim$10$^{32}$ erg$s^{-1}$). They also undergo occasional outbursts, when their luminosity increases by several orders of magnitude. During these events, BHTs can approach their Eddington luminosity and marked changes are observed in both time variability and energy spectrum (see e.g. \citealt{Belloni2005}, \citealt{Belloni2010}). 
We do not still have a complete understanding of all the mechanisms that lead to these changes, but, apart from the variation of the mass accretion rate, they must involve the structure of the accretion flow around the black hole as well as accretion/ejection mechanism processes. 

The spectral evolution of black hole X-ray transients can be described in terms of the characteristic pattern they usually show in the in X-ray hardness-intensity diagram (HID) (see  \citealt{Homan2001}, \citealt{Homan2005a}, \citealt{Belloni2006}, \citealt{Gierli'nski2006}, \citealt{Belloni2010}). 
Different states correspond to different branches/areas of the HID, which is often travelled along a regular path during outbursts. 
Two of the states correspond to the original states discovered in the 1970s. The Low/Hard State (LHS) is found only at the beginning and at the end of an outburst, where the highest luminosity swings are observed. The X-ray spectrum is dominated by a component which can be approximated by a power law with a hard photon index ($\sim$ 1.5-1.8) and a variable high-energy cutoff moving between $\sim$50 keV and $\sim$300 keV (see e.g. \citealt{Motta2009}, \citealt{Joinet2008}, \citealt{Miyakawa2008}). In this phase, the power density spectrum (PDS) of the source is highly variable and is dominated by a strong band limited noise ($\sim30\%$ fractional rms). The High Soft State (HSS), if reached, can be observed in the central part of an outburst, in which the spectrum is dominated by a soft thermal component, most likely associated to an optically thin accretion disk. It also shows an additional weak, steep power-law component (photon index $\sim$ 2.5 or higher). During this state, the power density spectrum shows faint components with fractional rms typically around few percent. 

Between these two well-established states, the situation is more complex, leading to many different classifications. \cite{Homan2005a} identify two additional states, defined by
spectral/timing transitions. After the LHS a transition to these intermediate states occurs and the source evolves as its luminosity increases. The spectrum starts to change: the soft thermal component appears and becomes gradually important, the energy peak of the emission softens and the hard component steepens ($\Gamma\sim$ 2.0-2.5). The Hard Intermediate State (HIMS) and Soft Intermediate State (SIMS) show these spectral characteristics and can be distinguished between each other mostly by timing properties (\citealt{Homan2005a}). The transition between HIMS and SIMS can be very fast (sometimes over a few seconds, see \citealt{Nespoli2003}) and it is marked by the disappearance/appearance of particular features in the PDS, such as the switch between two types of QPOs and a decrease in overall fast variability. The transition to the SIMS is also associated to the ejection of fast relativistic jets. This has led to the identification of a
\emph{jet line} in the HID, separating HIMS and SIMS (\citealt{Fender2004}). However, recent 
studies of different systems have shown that the jet ejection and HIMS/SIMS state transitions 
are not exactly simultaneous (\citealt{Fender2009}). The jet line can be crossed more than once
during an outburst (e.g. XTE J1859+226: \citealt{Casella2004}; GX339-4: \citealt{Brocksopp2002}, \citealt{Motta2009}). For a more detailed  state classification see \cite{Belloni2010}.

The four states mentioned above are observed in many BHCs in a regular way, starting from the LHS, crossing the HIMS and the SIMS and reaching the HSS. After a relatively long permanence in the HSS, the flux starts to decrease, most likely following a decrease in accretion rate. At some point, a reverse transition takes place and the path is followed backwards to the LHS and then to quiescence. The luminosity of this back-transition is always lower than that of the corresponding forward-transition (see \citealt{Maccarone2003}; \citealt{Dunn2010}). This basic pattern can vary depending on the source. Additional transitions between SIMS and HIMS can be observed and some sources behave in a more complicated way showing additional non-canonical states (i.e. the anomalous state, see \citealt{Belloni2010}).
Interestingly, until now all black-hole transients have shown two types of behaviour: after the initial LHS, most sources show a transition to the HIMS at a luminosity level which is always different and might be related to the previous history of the transient (\citealt{Yu2007}). If this transition takes place, the source always reached the HSS.
There is also a second group of systems (at least seven) that never left the LHS (eg. XTE1550-564, \cite{Sturner2005};).
%as reported for example by \cite{Brocksopp2004} for V404 Cyg, A1524-62, 4U1543-475, GRO J0422+32, GRO J1719-24, GRS1737-21 and GS 1354-64; by \cite{Sturner2005} for XTE1550-564 and by \cite{Rodriguez2006} for Aql X-1. 
Until the October 2008 outburst of H1743-322 (see below) the only possible exception to this dichotomy is represented by SAX~J1711.6--38 (\citealt{Wijnands2002}), a faint transient X-ray binary classified as black hole candidate. 

\subsection{H1743-322}

The X-ray source H1743-322 was discovered during a bright outburst in 1977 with the Ariel V satellite (\citealt{Kaluzienski1977}). 
It is classified as a black hole candidate (\citealt{McClintock2006}), as a dynamical confirmation has not been possible due to the faintness of its optical counterpart (\citealt{Steeghs2003}). 
H1743-322 is one of the few sources where X--ray jets have been imaged (\citealt{Corbel2005}). 
%The other sources are the BHTs 4U 1755−33 and XTE J1550--564 (\citealt{Corbel2002}; \citealt{Angelini2003}), the neutron star source Cir X-1 (\citealt{Heinz2007}; \citealt{Soleri2009}) and the peculiar source SS 433 (\citealt{Migliari2002}). 

The distance to H1743-322 is not well constrained. \cite{Corbel2005} find from the proper motion of the jet an upper limit of 10.4$\pm$2.9 kpc to the distance, consistent with a location in the Galactic Centre. 

 After displaying four outbursts (observed by different missions, such as INTEGRAL, {\it Swift}, RXTE), on 2008 
%In 2003, a new bright outburst was detected by INTEGRAL and the source was then dubbed IGR J17464-3213. Soon after it was realized that it corresponded to H1743-322 (\citealt{Markwardt2003}). After this outburst, H1743-322 underwent two fainter outbursts on September 2004 and September 2005 (\citealt{Swank2004}; \citealt{Rupen2005}; \citealt{Capitanio2006}). on Jenuary another outburst was detected by RXTE/ASM (\citealt{Kalemci2008}). Seven months later (on 
September 23, (MJD=54732), another outburst of H1743-322 was detected by INTEGRAL during the Galactic bulge monitoring (see Kuulkers et al. 2008)  showing that the source was in a hard state with an increasing flux. {\it Swift}, RXTE and INTEGRAL followed the outburst evolution. This time H1743-322 did not follow the canonical pattern normally seen in the outburst phases of black-hole transients. The source only sampled the HIMS (\citealt{Belloni2008}). Then it decreased in luminosity and underwent a hardening of the spectrum. This behaviour was confirmed by both spectral and timing analysis (\citealt{Capitanio2009}, hereafter C09). Finally, a new bright outburst was detected on May 2009 by Swift/BAT (see \citealt{Krimm2009}). This outburst followed the standard pattern in the HID,  displaying  all the canonical states.

In this paper we present the results coming from the analysis of the last two outbursts (2008 and 2009).
Our aim is to use RXTE data collected during the 2009
outburst to study the broadband spectral evolution of H1743-322 and compare its behavior with that of the 2008 outburst where a non-standard behavior is observed. Our purpose was to identify differences in the evolution that would have allowed to predict the presence/absence of a transition to soft states. This is important because the transition is linked to jet ejection. In 2008 the mechanism for such ejection was put in motion, but stopped early. In 2009 the transition took place. 
%----------------------------------------------------------------------------------
\begin{figure*}
\centering
\includegraphics[width=17.5cm]{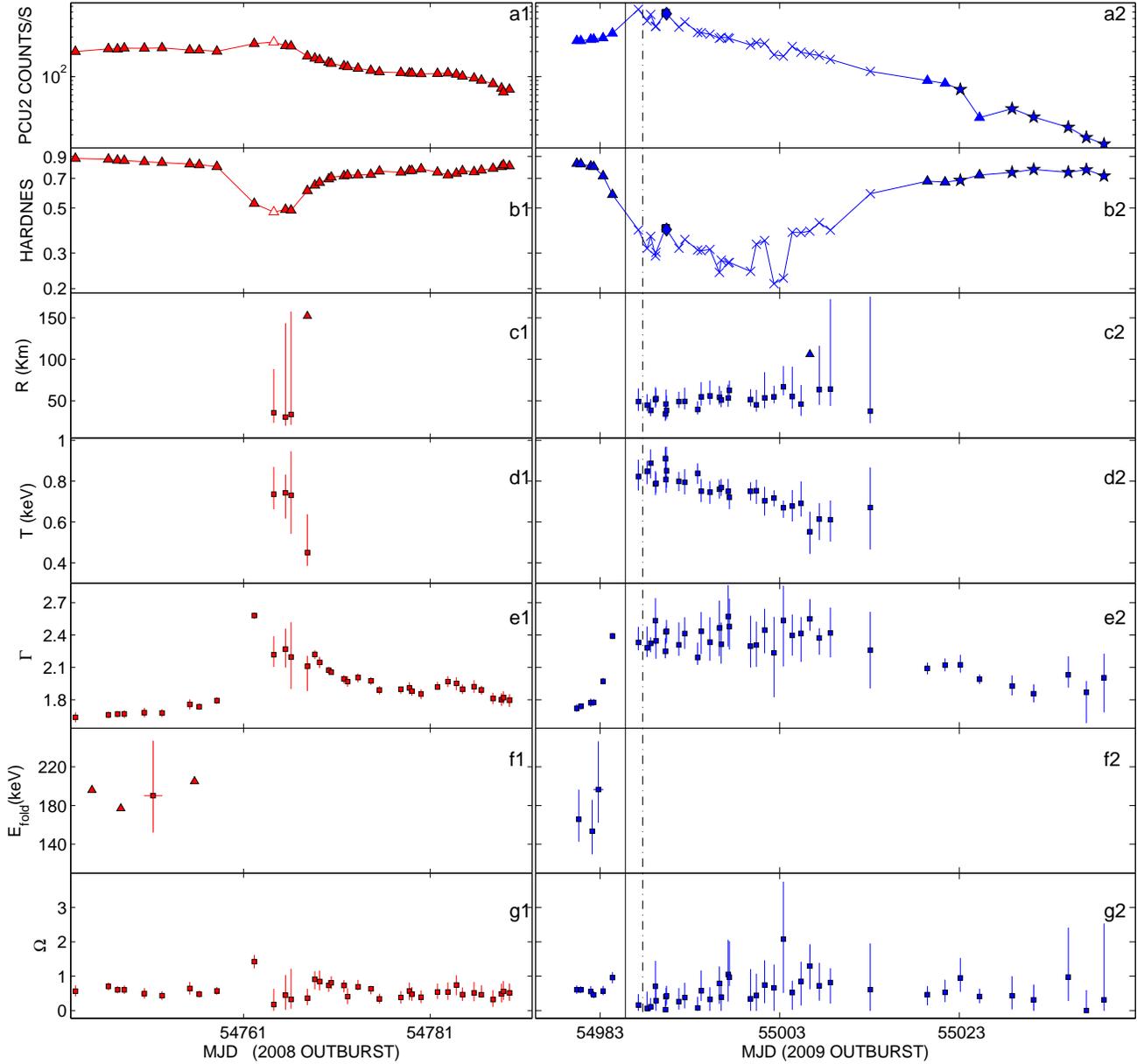}
\caption{Evolution of net count rate, hardness and main spectral parameters during the 2008 (red) and 2009 (blue) outburst of H1743-322 (see Tabs. \ref{tab:spettrali_2009} and \ref{tab:spettrali_2008}. 
From top to bottom: net PCU2 count rate, hardness ratio, inner disk radius in km (assuming a distance of 10 kpc and inclination of 65 degrees), disk temperature in keV at the inner disk radius, photon index, cutoff energy in keV, reflection factor. The solid line marks the transition from the HIMS to the SIMS and the dot-dashed line separates the SIMS from the HSS. Notice that only the primary transition from HIMS to SIMS is marked. The same transition takes place also later in the outburst (see text). No such transitions were observed in 2008. Points with horizontal error bars correspond to spectra obtained averaging observations with similar hardness: the error bars represent the time interval corresponding to the accumulation. In panel f, we used the values coming from Table \ref{tab:somme}. In panels a1, a2 and b1, b2 different symbols indicate different timing properties: type-A QPOs (diamonds), type-B QPOs (squares), type-C QPOs (filled triangles), strong band-limited noise components in the Power Density Spectrum (stars), weak band-limited noise components in the Power Density Spectrum (empty triangles), weak powerlaw noise in the Power Density Spectrum (crosses) Triangles in panels c1, c2 and f1, f2 mark upper limits.}
\label{fig:parametri}
\end{figure*}
%----------------------------------------------------------------------------------
\section{Observations and data analysis}\label{sec:observations} 
In May 2009, Swift/BAT detected X-ray activity from H1743-322 as part of its hard X-ray transient monitor program (\citealt{Krimm2009}).
This flux level was comparable to the peak reached during the previous outburst of this source around 4-October-2008 (see e.g. Kuulkers et al. 2008).
The brightening of H1743-322 was confirmed by the RXTE/ASM (\citealt{Miller-Jones2009}) during May 23-28. 
As a part of public target of opportunity observations H1743-322 was observed in pointing mode by RXTE. Starting from 29 May 2009, a total of 44 observations were performed during 2 months, covering a large part of the outburst of the source. We report here the results of the spectral analysis of all the 2009 observations. We also analysed the 37 RXTE outburst observations taken during 2008.

We extracted energy spectra from the PCA and HEXTE instruments (background and dead time
corrected) for each observation using the standard RXTE software
within HEASOFT V. 6.6.3. For our spectral analysis, only Proportional Counter Unit 2 from the PCA and
Cluster B from HEXTE were used. A systematic error of $0.6\%$
was added to the PCA spectra to account for residual
uncertainties in the instrument calibration \footnote{See http://www.universe.nasa.gov/xrays/programs/rxte/pca/doc/rmf/pcarmf-11.7/\#head-27884cefb2a102a7e53547f1631cbeab44224a04 for a detailed discussion on the PCA calibration issues.}. We accumulated
background corrected PCU2 rates in the channel bands A = 4 - 128
(3.3 - 118 keV), B = 4 - 10 (3.3 - 6.1 keV) and C = 11 - 20 (6.1
- 10.2 keV). A is the total count rate, while the hardness was defined
as H = C/B (\citealt{Homan2005a}). PCA+HEXTE spectra were fitted with \textsc{XSPEC V. 11} in the energy range 3 - 22 keV and 20-200 keV respectively. To account for cross-calibration problems, a variable multiplicative constant for the HEXTE spectra (as compared to the PCA) was added to the fits. 

For our timing analysis, we used custom software under
\textsc{IDL}. For each observation we produced power density spectra (PDS) from stretches
16 seconds long in the channel band 0-35 (2-15 keV). We averaged the PDS and subtracted the
contribution due to Poissonian noise (see \citealt{Zhang1995}) to produce a PDS for each
observation. They were normalized according to \cite{Leahy1983} and converted to squared fractional rms (\citealt{Belloni1990}). The integrated fractional rms was calculated over the 0.1 - 64 Hz band. 

\section{Results}\label{sec:result}

\subsection{The 2009 outburst of H1743-322}
\begin{figure*}
\begin{center}
\includegraphics[width=17.5cm]{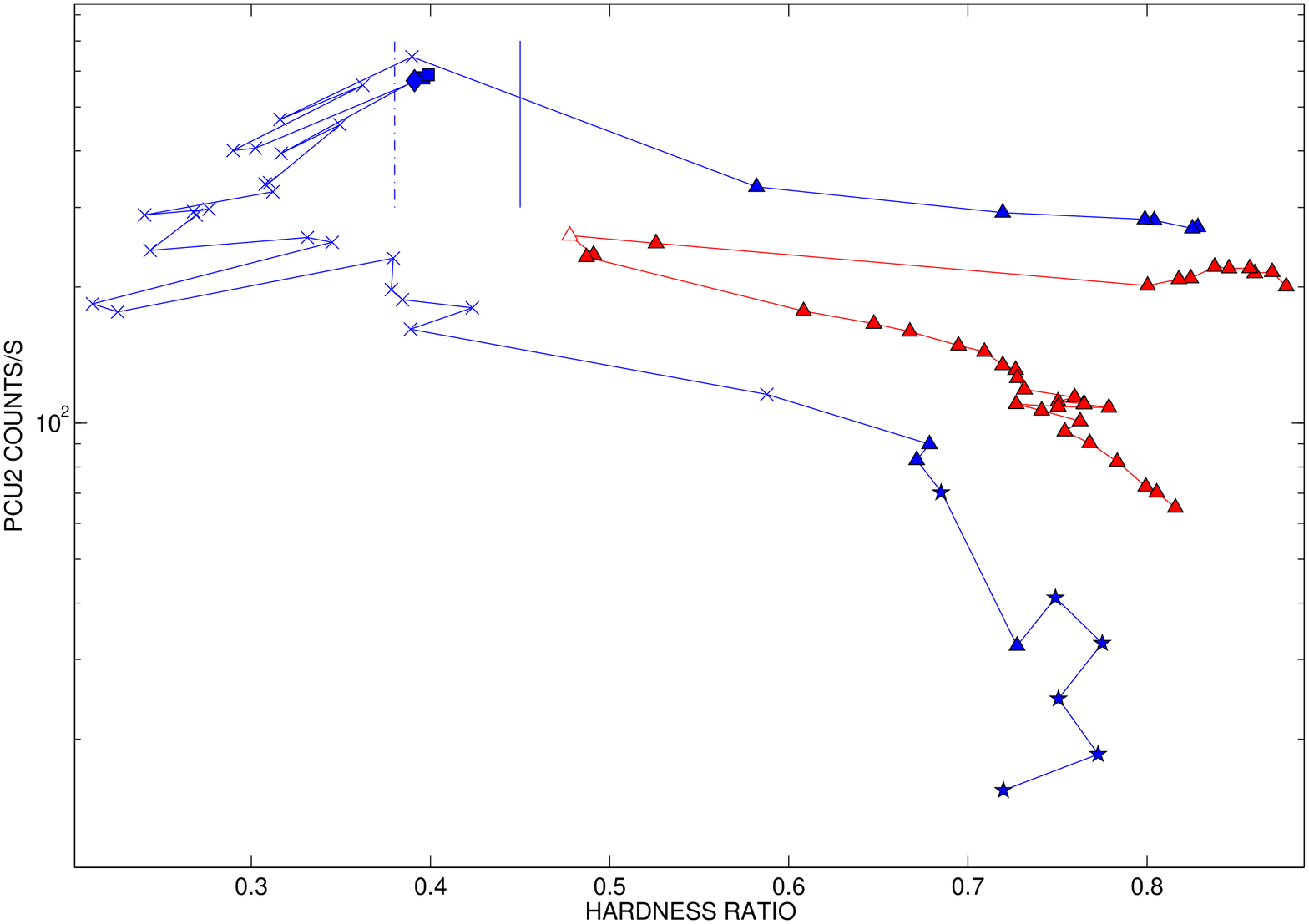}
\caption{Hardness-Intensity diagram from RXTE/PCA data for the complete 2009 outburst (blue line) and 2008 outburst (red line). The two paths start from the upper right corner and proceed in a counter-clockwise direction. Different symbols indicate different timing properties: type-A QPOs (diamonds), type-B QPOs (squares), type-C QPOs (filled triangles), strong band-limited noise components in the Power Density Spectrum (stars), weak band-limited noise components in the Power Density Spectrum (empty triangles), weak powerlaw noise in the Power Density Spectrum (crosses). The solid line marks thetransition from the HIMS to the SIMS and the dot-dashed line separates the SIMS from the HSS.}\label{fig:HID}
\end{center}
\end{figure*}

In this section, we describe the general evolution of the outburst. In Fig. \ref{fig:parametri} (panels a2 and b2) we show the count rate and hardness evolution of H1743-322 and in Fig. \ref{fig:HID} we show the HID. In Tab.\ref{tab:states_2009} we list the background-corrected PCU2 count rate, the hardness ratio and the fractional rms for each observation. 

%----------------------------------------------------------------------------------
%----------------------------------------------------------------------------------

As one can see from Fig. \ref{fig:HID} (blue track), the source evolution during the 2009 outburst is consistent with the typical behavior observed in most other BHTs (see e.g. \citealt{Gierli'nski2006}, \citealt{Remillard2006}, \citealt{Belloni2010}, \citealt{Fender2009}). Unfortunately the data did not cover the whole evolution from quiescence and the right branch of the outburst (i.e. LHS) was missed. However, the source was followed during its evolution from the HIMS through all the remaining canonical states. We observed a clear horizontal branch in the HID characterized by a sligtly rising  count rate and a progressive softening which drove the source from hard to soft state in few days. After a relatively long permanence in the soft state ( $\sim$ 22 days), the transition from the soft back to the hard state took place at a lower count rate than the initial opposite transition, as expected. 

%%_____________________BEGIN__________TABLE 1____________________________%%
\begin{table*} 
\renewcommand{\arraystretch}{1.3} 
\begin{center} 
\begin{tabular}{|c|c|c|c|c|c|c|c|} 
\multicolumn{8}{|c|}{2009 Outburst} \\
\hline 
Obs no. & obs ID & MJD & color & PCU2 Counts s$^{}$ & rms & QPO & State \\
\hline 
1 & 94413-01-02-00 & 54980.40 & $ 0.828 \pm 0.004 $ & $ 271.2 \pm 0.6 $ & $ 30.3 \pm 0.4 $ & C & HIMS \\
2 & 94413-01-02-02 & 54980.84 & $ 0.825 \pm 0.003 $ & $ 269.5 \pm 0.4 $ & $ 29.7 \pm 0.4 $ & C & HIMS \\
3 & 94413-01-02-01 & 54981.95 & $ 0.804 \pm 0.004 $ & $ 280.9 \pm 0.6 $ & $ 29.7 \pm 0.4 $ & C & HIMS \\
4 & 94413-01-02-05 & 54982.28 & $ 0.799 \pm 0.004 $ & $ 282.2 \pm 0.5 $ & $ 30.1 \pm 0.3 $ & C & HIMS \\
5 & 94413-01-02-04 & 54983.33 & $ 0.719 \pm 0.003 $ & $ 291.9 \pm 0.5 $ & $ 27.7 \pm 0.3 $ & C & HIMS \\
6 & 94413-01-02-03 & 54984.37 & $ 0.582 \pm 0.002 $ & $ 333.1 \pm 0.6 $ & $ 22.3 \pm 0.1 $ & C & HIMS \\
\hline 
7 & 94413-01-03-00 & 54987.26 & $ 0.390 \pm 0.002 $ & $ 645.3 \pm 1.2 $ & $ 4.5 \pm 0.9 $ & - & SIMS \\
\hline 
8 & 94413-01-03-01 & 54988.23 & $ 0.316 \pm 0.001 $ & $ 469.6 \pm 0.9 $ & $ 5.1 \pm 0.5 $ & - & HSS \\
9 & 94413-01-03-07 & 54988.63 & $ 0.362 \pm 0.001 $ & $ 558.2 \pm 1.0 $ & $ 3.4 \pm 0.4 $ & - & HSS \\
10 & 94413-01-03-05 & 54989.15 & $ 0.290 \pm 0.002 $ & $ 400.8 \pm 0.9 $ & $ 4.6 \pm 0.7 $ & - & HSS \\
11 & 94413-01-03-06 & 54989.22 & $ 0.302 \pm 0.002 $ & $ 405.6 \pm 0.9 $ & $ 5.7 \pm 0.5 $ & - & HSS \\
\hline 
12 & 94413-01-03-02 & 54990.26 & $ 0.396 \pm 0.002 $ & $ 580.1 \pm 1.1 $ & $ 4.9 \pm 0.9 $ & B & SIMS \\
13 & 94413-01-03-03 & 54990.33 & $ 0.399 \pm 0.002 $ & $ 589.3 \pm 1.1 $ & $ 6.5 \pm 0.6 $ & B & SIMS \\
14 & 94413-01-03-04 & 54990.39 & $ 0.391 \pm 0.002 $ & $ 571.6 \pm 1.0 $ & $ 3.4 \pm 0.3 $ & A & SIMS \\
\hline 
15 & 94413-01-03-08 & 54991.77 & $ 0.317 \pm 0.001 $ & $ 394.7 \pm 0.7 $ & $ 5.4 \pm 0.3 $ & - & HSS \\
16 & 94413-01-03-09 & 54992.43 & $ 0.349 \pm 0.002 $ & $ 456.7 \pm 0.9 $ & $ 3.2 \pm 0.7 $ & - & HSS \\
17 & 94413-01-03-10 & 54993.86 & $ 0.310 \pm 0.001 $ & $ 340.1 \pm 0.7 $ & $ 7.5 \pm 0.4 $ & - & HSS \\
18 & 94413-01-04-00 & 54994.25 & $ 0.308 \pm 0.002 $ & $ 337.6 \pm 0.8 $ & $ 6.0 \pm 1.2 $ & - & HSS \\
19 & 94413-01-04-01 & 54995.23 & $ 0.312 \pm 0.002 $ & $ 324.7 \pm 0.8 $ & $ 5.2 \pm 0.8 $ & - & HSS \\
20 & 94413-01-04-03 & 54996.28 & $ 0.240 \pm 0.001 $ & $ 288.7 \pm 0.7 $ & $ 4.8 \pm 1.7 $ & - & HSS \\
21 & 94413-01-04-02 & 54996.48 & $ 0.276 \pm 0.001 $ & $ 297.3 \pm 0.6 $ & $ 4.7 \pm 1.3 $ & - & HSS \\
22 & 94413-01-04-08 & 54997.26 & $ 0.268 \pm 0.002 $ & $ 293.5 \pm 0.7 $ & $ 3.7 \pm 1.2 $ & - & HSS \\
23 & 94413-01-04-05 & 54997.39 & $ 0.269 \pm 0.001 $ & $ 288.4 \pm 0.7 $ & $ 4.9 \pm 1.5 $ & - & HSS \\
24 & 94413-01-04-06 & 54999.75 & $ 0.243 \pm 0.001 $ & $ 241.0 \pm 0.6 $ & $ 3.6 \pm 1.3 $ & - & HSS \\
25 & 94413-01-04-04 & 55000.40 & $ 0.331 \pm 0.002 $ & $ 257.3 \pm 0.6 $ & $ 5.2 \pm 1.4 $ & - & HSS \\
26 & 94413-01-05-00 & 55001.32 & $ 0.345 \pm 0.002 $ & $ 251.1 \pm 0.6 $ & $ 7.0 \pm 0.6 $ & - & HSS \\
27 & 94413-01-05-01 & 55002.36 & $ 0.211 \pm 0.001 $ & $ 183.5 \pm 0.5 $ & $ 2.1 \pm 2.2 $ & - & HSS \\
28 & 94413-01-05-02 & 55003.41 & $ 0.225 \pm 0.001 $ & $ 176.2 \pm 0.4 $ & $ 1.7 \pm 2.6 $ & - & HSS \\
29 & 94413-01-05-03 & 55004.39 & $ 0.379 \pm 0.002 $ & $ 231.6 \pm 0.5 $ & $ 6.4 \pm 0.7 $ & - & HSS \\
30 & 94413-01-05-04 & 55005.37 & $ 0.378 \pm 0.002 $ & $ 197.3 \pm 0.5 $ & $ 7.5 \pm 0.6 $ & - & HSS \\
31 & 94413-01-05-05 & 55006.35 & $ 0.384 \pm 0.002 $ & $ 187.5 \pm 0.4 $ & $ 6.8 \pm 0.7 $ & - & HSS \\
32 & 94413-01-05-06 & 55007.40 & $ 0.423 \pm 0.002 $ & $ 179.9 \pm 0.4 $ & $ 9.4 \pm 0.4 $ & - & HSS \\
33 & 94413-01-06-00 & 55008.64 & $ 0.389 \pm 0.002 $ & $ 161.3 \pm 0.4 $ & $ 9.2 \pm 0.7 $ & - & HSS \\
34 & 94413-01-06-01 & 55013.09 & $ 0.588 \pm 0.005 $ & $ 115.7 \pm 0.4 $ & $ 12.9 \pm 1.6 $ & - & HSS \\
\hline 
35 & 94413-01-07-00 & 55016.32 & $ 1.266 \pm 0.011 $ & $ 205.6 \pm 0.8 $ & $ 16.6 \pm 1.6 $ & C & HIMS \\
36 & 94413-01-07-01 & 55019.45 & $ 0.679 \pm 0.006 $ & $ 89.9 \pm 0.3 $ & $ 19.8 \pm 0.4 $ & C & HIMS \\
37 & 94413-01-07-02 & 55021.42 & $ 0.671 \pm 0.006 $ & $ 82.9 \pm 0.4 $ & $ 18.8 \pm 0.5 $ & C & HIMS \\
38 & 94413-01-08-02 & 55023.11 & $ 0.685 \pm 0.008 $ & $ 70.2 \pm 0.4 $ & $ 18.5 \pm 2.9 $ & - & HIMS \\
39 & 94413-01-08-00 & 55025.25 & $ 0.727 \pm 0.005 $ & $ 32.2 \pm 0.1 $ & $ 18.1 \pm 0.5 $ & C & HIMS \\
\hline 
40 & 94413-01-08-01 & 55028.88 & $ 0.749 \pm 0.011 $ & $ 41.1 \pm 0.3 $ & $ 16.9 \pm 2.8 $ & - & LHS \\
41 & 94413-01-09-00 & 55031.29 & $ 0.775 \pm 0.011 $ & $ 32.6 \pm 0.3 $ & $ 18.6 \pm 2.3 $ & - & LHS \\
42 & 94413-01-09-01 & 55035.14 & $ 0.751 \pm 0.013 $ & $ 24.6 \pm 0.2 $ & $ 18.0 \pm 3.1 $ & - & LHS \\
43 & 94413-01-10-00 & 55037.16 & $ 0.773 \pm 0.014 $ & $ 18.5 \pm 0.2 $ & $ 15.5 \pm 9.0 $ & - & LHS \\
44 & 94413-01-10-01 & 55039.12 & $ 0.720 \pm 0.016 $ & $ 15.4 \pm 0.2 $ & $ 13.0 \pm 10.0 $ & - & LHS \\
\hline 
\end{tabular} 
\caption{The columns are: observation number for the 2009 outburst, RXTE observation ID, MJD, PCU2 count rate, hardness ratio, integrated fractional rms (0.1 - 64 Hz), low-frequency QPO type, and state according to Belloni (2010). }\label{tab:states_2009} 
\end{center} 
\end{table*} 
%%_____________________END__________TABLE 1____________________________%%
\subsubsection{Timing Analysis}\label{sec:timing}
%The evolution of a BHT is usually described in terms of spectral states. 
Source states are defined on the basis of both spectral and timing properties (for a detailed description of state classification see \citealt{Homan2005a}; \citealt{Belloni2005}; \citealt{Belloni2010}, \citealt{Fender2009}). Thus, timing information (i.e. PDS) is needed in order to identify the branches we see in the HID in terms of states. This will serve as a framework for the spectral analysis (see Sec. \ref{sec:spectral_analysis}).
In Tab. \ref{tab:timing_tab_2009} we outline the main conclusions extracted from the PDS analysis.
\begin{itemize}
\item Observations $\#1$ to $\#6$ and observation $\#35$ to $\#39$ show a high level of aperiodic variability in the form of strong band-limited noise component (flat top noise shape), with total integrated fractional rms in the range $17 - 30\%$. The rms is positively correlated with hardness. The PDS can be decomposed in a number of Lorentzian components, one of which takes the form of a type-C QPO peak (see \cite{Casella2004} for a complete QPO classification).
The observations correspond to the first part of the horizontal branch (Fig. \ref{fig:HID}, triangles). These observations belong to the HIMS.
\item Observation \#7 shows a weak variability level and a powerlaw shape noise, associated to a low fractional rms value ($\sim 5\%$). These values and the hardness are consistent with the values observed in observations $\#12$, $\#13$ and $\#14$, where the systems is in the SIMS (see below). No QPOs are observed here. However we note that Type-A QPOs are not always detected in PDS associated to the SIMS due to their faintness. We tentatively classify this observation as SIMS.
\item Observations $\#8$ to $\#11$ and $\#15$ to $\#34$ show weak powerlaw noise with a rms of a few percent. They correspond to the softest observations. All these observations are marked with crosses in the HID in Fig. \ref{fig:HID} and belongs to the HSS.
\item Observations $\#12$ and $\#13$ correspond to a low variability ($\sim 5-6\%$ fractional rms). A type-B QPO is prominent in the PDS, significanly different from the ones observed in the LHS and in the HIMS (see \citealt{Motta2009}, \citealt{Belloni2005}; \citealt{Belloni2008}). These observations are marked in the HID (Fig. \ref{fig:HID}) with diamonds and are classified as SIMS. 
\item Observation $\#14$ shows a Type-A QPO and has an integrated fractional rms of $\sim 3.4\%$. This observation is also softer than the two showing a Type-B QPO. This observations also belongs to the SIMS (in the middle of the HID, Fig. \ref{fig:HID}) and is marked in the HID with a square.
\item Observation from $\#40$ to $\#44$ show a quite high level of aperiodic variability in the form of strong band-limited noise components, with fractional rms in the range $13-18\%$. The PDS are similar to those observed during the HIMS and can be decomposed in a number of Lorentzian components. These observations correspond to the vertical branch of the HID in Fig. \ref{fig:HID} (stars), when the source undergoes the final hardening before the quiescence phase. These observations belong to the LHS.
\end{itemize}
%
%%_____________________BEGIN________TABLE 2____________________________%%
\begin{table*}
\renewcommand{\arraystretch}{1.3}
\begin{center}
\begin{tabular}{|c|c|c|c|}
\multicolumn{4}{|c|}{2009 Outburst} \\
\hline 
\#obs ID & Noise Type & QPO type & RMS (in \%) \\
\hline
\hline
\#1 to \#6, \#35 to \#39 & strong band limited noise & C & 17-30\% \\
\#7 & weak powerlaw component & - & 5\% \\
\#8 to \#11; \#15 to \#34 & weak powerlaw component & - & 3-12\% \\
\#12,\#13 & weak powerlaw component & B & 5-6\% \\
\#14 & weak powerlaw component & A & 3.4\% \\
\#40 to \#44 & strong band limited noise & C & 13-18\% \\
\hline
\end{tabular}
\caption{Summary of the timing properties seen in the PDS of each observation from the 2009 outburst. }\label{tab:timing_tab_2009}
\end{center}
\end{table*}
%%_____________________END________TABLE 2____________________________%%
On the basis of the state classification, we can identify three main transitions and one backward transition during the softening of the source. Here by we will call main transitions to the ones that take place in the HIMS-SIMS-HSS direction for the first time and secondary the following transitions in the same directions. Backward transitions take place in the inverse direction. 
\begin{itemize}
\item Main transition from HIMS to SIMS, taking place between observations $\#6$ and $\#7$ at hardness $\sim$ 0.4. 
\item Main transition from SIMS to HSS, taking place between observations $\#7$ and $\#8$ at hardness $\sim$ 0.35. 
\item A secondary transition from SIMS to HSS, taking place between Obs. \#14 and \#15, at hardness $\sim$ 0.35.
\item Backward transition from HSS to SIMS, taking place between observations $\#11$ and $\#12$ at hardness $\sim$ 0.3. As it is often observed in other sources (\citealt{Motta2009}, \citealt{DelSanto2009}) the SIMS is crossed several times during the outburst evolution, either in the HISM-SIMS-HSS direction or in the opposite one.
\end{itemize}
Lines corresponding to the main transitions are shown in Fig. 1 and Fig. 2.

%%_____________________BEGIN__________TABLE_3____________________________%%
\begin{table*} 
\renewcommand{\arraystretch}{1.3} 
\begin{center} 
\begin{tabular}{|c|c|c|c|c|c|} 
\multicolumn{6}{|c|}{2009 Outburst} \\
\hline 
obs no. & kT (keV) & R (Km) & $\Gamma$ & $E_f (keV)$ & $\Omega$ \\
\hline 
1 & $ - $ & $ - $ & $ 1.72 _{- 0.03 }^{+ 0.03 } $ & $ 148 _{- 29 }^{+ 43 } $ & $ 0.6 _{- 0.1 }^{+ 0.1 } $ \\
2 & $ - $ & $ - $ & $ 1.74 _{- 0.03 }^{+ 0.03 } $ & $ 170 _{- 29 }^{+ 42 } $ & $ 0.6 _{- 0.1 }^{+ 0.1 } $ \\
3 & $ - $ & $ - $ & $ 1.77 _{- 0.03 }^{+ 0.04 } $ & $ 156 _{- 37 }^{+ 62 } $ & $ 0.6 _{- 0.1 }^{+ 0.1 } $ \\
4 & $ - $ & $ - $ & $ 1.78 _{- 0.03 }^{+ 0.03 } $ & $ 164 _{- 29 }^{+ 55 } $ & $ 0.5 _{- 0.1 }^{+ 0.1 } $ \\
5 & $ - $ & $ - $ & $ 1.97 _{- 0.03 }^{+ 0.03 } $ & $ > 200 $ & $ 0.6 _{- 0.1 }^{+ 0.1 } $ \\
6 & $ - $ & $ - $ & $ 2.39 _{- 0.03 }^{+ 0.02 } $ & $ > 200 $ & $ 1.0 _{- 0.2 }^{+ 0.1 } $ \\
7 & $ 0.82 _{- 0.07 }^{+ 0.08 } $ & $ 49 _{- 11 }^{+ 16 } $ & $ 2.33 _{- 0.07 }^{+ 0.14 } $ & $ > 200 $ & $ 0.2 _{- 0.2 }^{+ 0.3 } $ \\
8 & $ 0.85 _{- 0.06 }^{+ 0.05 } $ & $ 45 _{- 7 }^{+ 13 } $ & $ 2.28 _{- 0.08 }^{+ 0.20 } $ & $ > 200 $ & $ 0.1 _{- 0.1 }^{+ 0.5 } $ \\
9 & $ 0.89 _{- 0.07 }^{+ 0.07 } $ & $ 38 _{- 7 }^{+ 14 } $ & $ 2.32 _{- 0.08 }^{+ 0.05 } $ & $ > 200 $ & $ 0.1 _{- 0.1 }^{+ 0.2 } $ \\
10 & $ 0.79 _{- 0.05 }^{+ 0.06 } $ & $ 52 _{- 10 }^{+ 14 } $ & $ 2.53 _{- 0.22 }^{+ 0.20 } $ & $ > 200 $ & $ 0.7 _{- 0.5 }^{+ 0.7 } $ \\
11 & $ 0.79 _{- 0.04 }^{+ 0.05 } $ & $ 53 _{- 9 }^{+ 12 } $ & $ 2.35 _{- 0.16 }^{+ 0.17 } $ & $ > 200 $ & $ 0.3 _{- 0.3 }^{+ 0.5 } $ \\
12 & $ 0.91 _{- 0.11 }^{+ 0.05 } $ & $ 34 _{- 8 }^{+ 14 } $ & $ 2.25 _{- 0.06 }^{+ 0.12 } $ & $ > 200 $ & $ 0.0 _{- 0.0 }^{+ 0.3 } $ \\
13 & $ 0.81 _{- 0.06 }^{+ 0.10 } $ & $ 46 _{- 11 }^{+ 17 } $ & $ 2.43 _{- 0.11 }^{+ 0.05 } $ & $ > 200 $ & $ 0.4 _{- 0.3 }^{+ 0.3 } $ \\
14 & $ 0.85 _{- 0.08 }^{+ 0.12 } $ & $ 38 _{- 10 }^{+ 18 } $ & $ 2.43 _{- 0.10 }^{+ 0.10 } $ & $ > 200 $ & $ 0.4 _{- 0.2 }^{+ 0.3 } $ \\
15 & $ 0.80 _{- 0.05 }^{+ 0.04 } $ & $ 49 _{- 7 }^{+ 11 } $ & $ 2.31 _{- 0.10 }^{+ 0.20 } $ & $ > 200 $ & $ 0.3 _{- 0.2 }^{+ 0.4 } $ \\
16 & $ 0.79 _{- 0.06 }^{+ 0.06 } $ & $ 49 _{- 9 }^{+ 16 } $ & $ 2.41 _{- 0.14 }^{+ 0.15 } $ & $ > 200 $ & $ 0.4 _{- 0.3 }^{+ 0.4 } $ \\
17 & $ 0.84 _{- 0.05 }^{+ 0.05 } $ & $ 39 _{- 6 }^{+ 10 } $ & $ 2.19 _{- 0.07 }^{+ 0.14 } $ & $ > 200 $ & $ 0.1 _{- 0.1 }^{+ 0.3 } $ \\
18 & $ 0.75 _{- 0.05 }^{+ 0.06 } $ & $ 55 _{- 10 }^{+ 17 } $ & $ 2.43 _{- 0.21 }^{+ 0.18 } $ & $ > 200 $ & $ 0.6 _{- 0.3 }^{+ 0.6 } $ \\
19 & $ 0.75 _{- 0.06 }^{+ 0.05 } $ & $ 56 _{- 10 }^{+ 18 } $ & $ 2.33 _{- 0.17 }^{+ 0.23 } $ & $ > 200 $ & $ 0.3 _{- 0.3 }^{+ 0.3 } $ \\
20 & $ 0.76 _{- 0.05 }^{+ 0.04 } $ & $ 54 _{- 8 }^{+ 13 } $ & $ 2.46 _{- 0.26 }^{+ 0.25 } $ & $ > 200 $ & $ 0.8 _{- 0.6 }^{+ 0.5 } $ \\
21 & $ 0.77 _{- 0.04 }^{+ 0.04 } $ & $ 51 _{- 8 }^{+ 11 } $ & $ 2.31 _{- 0.18 }^{+ 0.20 } $ & $ > 200 $ & $ 0.4 _{- 0.4 }^{+ 0.3 } $ \\
22 & $ 0.75 _{- 0.03 }^{+ 0.06 } $ & $ 53 _{- 10 }^{+ 16 } $ & $ 2.57 _{- 0.27 }^{+ 0.29 } $ & $ > 200 $ & $ 1.1 _{- 0.8 }^{+ 1.0 } $ \\
23 & $ 0.72 _{- 0.06 }^{+ 0.04 } $ & $ 63 _{- 9 }^{+ 11 } $ & $ 2.48 _{- 0.21 }^{+ 0.26 } $ & $ > 200 $ & $ 1.0 _{- 0.2 }^{+ 1.0 } $ \\
24 & $ 0.75 _{- 0.04 }^{+ 0.04 } $ & $ 52 _{- 8 }^{+ 12 } $ & $ 2.30 _{- 0.20 }^{+ 0.28 } $ & $ > 200 $ & $ 0.3 _{- 0.3 }^{+ 0.9 } $ \\
25 & $ 0.75 _{- 0.06 }^{+ 0.05 } $ & $ 45 _{- 8 }^{+ 18 } $ & $ 2.31 _{- 0.20 }^{+ 0.21 } $ & $ > 200 $ & $ 0.4 _{- 0.4 }^{+ 0.6 } $ \\
26 & $ 0.70 _{- 0.07 }^{+ 0.07 } $ & $ 54 _{- 12 }^{+ 31 } $ & $ 2.45 _{- 0.21 }^{+ 0.19 } $ & $ > 200 $ & $ 0.7 _{- 0.5 }^{+ 0.7 } $ \\
27 & $ 0.72 _{- 0.04 }^{+ 0.04 } $ & $ 55 _{- 8 }^{+ 13 } $ & $ 2.23 _{- 0.41 }^{+ 0.33 } $ & $ > 200 $ & $ 0.7 _{- 0.7 }^{+ 0.7 } $ \\
28 & $ 0.67 _{- 0.05 }^{+ 0.03 } $ & $ 67 _{- 10 }^{+ 25 } $ & $ 2.54 _{- 0.42 }^{+ 0.32 } $ & $ > 200 $ & $ 2.1 _{- 1.6 }^{+ 1.7 } $ \\
29 & $ 0.68 _{- 0.07 }^{+ 0.08 } $ & $ 55 _{- 14 }^{+ 36 } $ & $ 2.40 _{- 0.18 }^{+ 0.19 } $ & $ > 200 $ & $ 0.5 _{- 0.5 }^{+ 0.3 } $ \\
30 & $ 0.69 _{- 0.06 }^{+ 0.11 } $ & $ 46 _{- 14 }^{+ 23 } $ & $ 2.42 _{- 0.26 }^{+ 0.15 } $ & $ > 200 $ & $ 0.9 _{- 0.7 }^{+ 0.6 } $ \\
31 & $ 0.55 _{- 0.11 }^{+ 0.10 } $ & $ 106 _{- 36 }^{+ 318 } $ & $ 2.55 _{- 0.11 }^{+ 0.18 } $ & $ > 200 $ & $ 1.3 _{- 0.7 }^{+ 0.6 } $ \\
32 & $ 0.61 _{- 0.10 }^{+ 0.08 } $ & $ 63 _{- 18 }^{+ 52 } $ & $ 2.37 _{- 0.15 }^{+ 0.09 } $ & $ > 200 $ & $ 0.7 _{- 0.4 }^{+ 0.7 } $ \\
33 & $ 0.61 _{- 0.11 }^{+ 0.09 } $ & $ 64 _{- 20 }^{+ 108 } $ & $ 2.42 _{- 0.22 }^{+ 0.23 } $ & $ > 200 $ & $ 0.8 _{- 0.6 }^{+ 0.4 } $ \\
34 & $ 0.67 _{- 0.20 }^{+ 0.20 } $ & $ 37 _{- 14 }^{+ 138 } $ & $ 2.26 _{- 0.35 }^{+ 0.35 } $ & $ > 200 $ & $ 0.6 _{- 0.6 }^{+ 1.3 } $ \\
35 & $ - $ & $ - $ & $ 2.36 _{- 0.07 }^{+ 0.06 } $ & $ > 200 $ & $ 0.9 _{- 0.4 }^{+ 0.4 } $ \\
36 & $ - $ & $ - $ & $ 2.09 _{- 0.05 }^{+ 0.05 } $ & $ > 200 $ & $ 0.5 _{- 0.3 }^{+ 0.2 } $ \\
37 & $ - $ & $ - $ & $ 2.12 _{- 0.05 }^{+ 0.06 } $ & $ > 200 $ & $ 0.5 _{- 0.3 }^{+ 0.4 } $ \\
38 & $ - $ & $ - $ & $ 2.12 _{- 0.07 }^{+ 0.09 } $ & $ > 200 $ & $ 0.9 _{- 0.4 }^{+ 0.6 } $ \\
39 & $ - $ & $ - $ & $ 1.99 _{- 0.04 }^{+ 0.04 } $ & $ > 200 $ & $ 0.4 _{- 0.2 }^{+ 0.2 } $ \\
40 & $ - $ & $ - $ & $ 1.93 _{- 0.08 }^{+ 0.10 } $ & $ > 200 $ & $ 0.4 _{- 0.4 }^{+ 0.6 } $ \\
41 & $ - $ & $ - $ & $ 1.86 _{- 0.08 }^{+ 0.08 } $ & $ > 200 $ & $ 0.3 _{- 0.3 }^{+ 0.4 } $ \\
42 & $ - $ & $ - $ & $ 2.03 _{- 0.11 }^{+ 0.17 } $ & $ > 200 $ & $ 1.0 _{- 0.7 }^{+ 1.4 } $ \\
43 & $ - $ & $ - $ & $ 1.87 _{- 0.28 }^{+ 0.10 } $ & $ > 200 $ & $ 0.0 _{- 0.0 }^{+ 0.6 } $ \\
44 & $ - $ & $ - $ & $ 2.01 _{- 0.32 }^{+ 0.22 } $ & $ > 200 $ & $ 0.3 _{- 0.3 }^{+ 2.2 } $ \\
\hline 
\end{tabular}
\caption{Spectral parameters for the 2009 outburst of H1743-322. Columns are: observation number, inner disc temperature kT, inner disc radius R (assuming a distance of 10 kpc and an inclination of 65 $^o$), photon index $\Gamma$, fold Energy E$_{fold}$ (corresponding to high energy cutoff), reflection factor $\Omega$.}\label{tab:spettrali_2009} 
\end{center} 
\end{table*} 
%%_____________________END__________TABLE_3____________________________%%

\subsubsection{Spectral Analysis}\label{sec:spectral_analysis}

Given the multitude of spectral models available, we tested several ones in a first approach. This method has been already adopted for other sources (GX 339-4, \citealt{Nowak2002}; Cyg X-1, \citealt{Wilms2005}).
We started with a model consisting of only one component, either a cutoff power law or a multicolor disk blackbody, but neither could fit the spectra. Better results were obtained with a combination of these two components. However, this model does not yield good fits for all the observations: clear residuals around 30 keV are observed for some cases. They appear at the beginning and at the end of the outburst and we interpreted them as evidence of reflection of the Comptonized emission on relatively cold matter (i.e. the accretion disk). 
Moreover, during the initial and final periods of the outburst, the disk component has normalization values too low to be physically acceptable\footnote{Assuming a 10M$_\odot$ black hole, the  gravitational radius should be  $\sim$ 30 km. The normalizations found range between $\sim$5 and $\sim$230, leading to inner disk radii ranging between $>$3-19 km (assuming an inclination angle $<$ 65$^o$). This means that the maximum inner disk radius found is 0.62 gravitational radii, too small from a physical point of view.}. This model was used by C09 to fit the spectra corresponding to the 2008 outburst.
The model is not able to fit the observations that do not present a disk blackbody component. As mentioned above, an additional reflection component was required in order to get acceptable fits of the spectra without a visible disk component. Considering that, if a disk component is observable, the reflection component is expected to be proportional to the disk flux and correlated with the photon index (see \citealt{Gilfanov2009} and reference therein), we concluded that the model is not able to describe the spectrum in a physical way. %Therefore we choose to use a more physical spectral model that could describe all the observations, consistently with the expected physics. 

Following \cite{Wilms2006}, \cite{Nowak2005} and \citealt{Munoz-Darias2010}, we tried to model the data with an absorbed broken (elbow-shaped) power law associated to an exponentially high-energy cutoff. 
Even though the broken power-law shape mimics the effect of reflection, which is clearly seen in this source (see below), the fits suffered the same problems than using the cutoff powerlaw+diskblackbody model. We conclude that they are due to an oversimplification of the model itself, which is not sufficient to give a valid description of the spectra possibly because of the complex reflection features observed.

In order to obtain good fits and acceptable physical parameters, a model consisting of an exponentially cut off power law spectrum reflected from neutral material (\citealt{Magdziarz1995}) was used ({\tt pexrav} \footnote{The output spectrum consists of a power law with a high-energy cutoff combined with reflection from a neutral medium. $\Omega/ 2 \pi$ represents the fraction of the hard X-ray radiation emitted towards the disc where it is reflected.} in Xspec). An gaussian emission line with centroid fixed at 6.4 keV was further needed in order to obtain acceptable fits. The iron width was constrained between 0.2 and 1.2 keV to prevent artificial broadening due to the bad response of XTE/PCA at 6.4 keV. The use of {\tt pexrav} is justified by the presence of the iron line in all the spectra. In addition, a multi-color disk-blackbody ({\tt diskbb}) was added to the model.
A hydrogen column density was used ({\tt wabs} in XSPEC), with N$_{\rm H}$ frozen to $1.6 \times 10^{22} {\rm cm}^{-2}$, the value derived from {\it Swift}/XRT (C09). 

We tried to further improve the fits by adding an iron edge component to the model. This is justified by te fact that an iron edge is expected when the iron line is visible. The iron edge component was added with energy constrained between 6 and 10 keV and free optical depth. Although the $\chi^{2}_{\mathrm{red}}$ was generally slightly lower for most of the observations, the absorption edge was not well constrained. It tends to move to high energies, reaching the higher limit we imposed. In addition, the component is often not clearly significant and the optical depth obtained is usually lower than $\sim$ 0.1. We then concluded that the addition of the iron absorption edge is not justified. Even though we cannot exclude the presence of the iron edge component, we can assert that, if present, it is too faint to be clearly observed and constrained. %

\begin{figure}
\includegraphics[width=8.5cm]{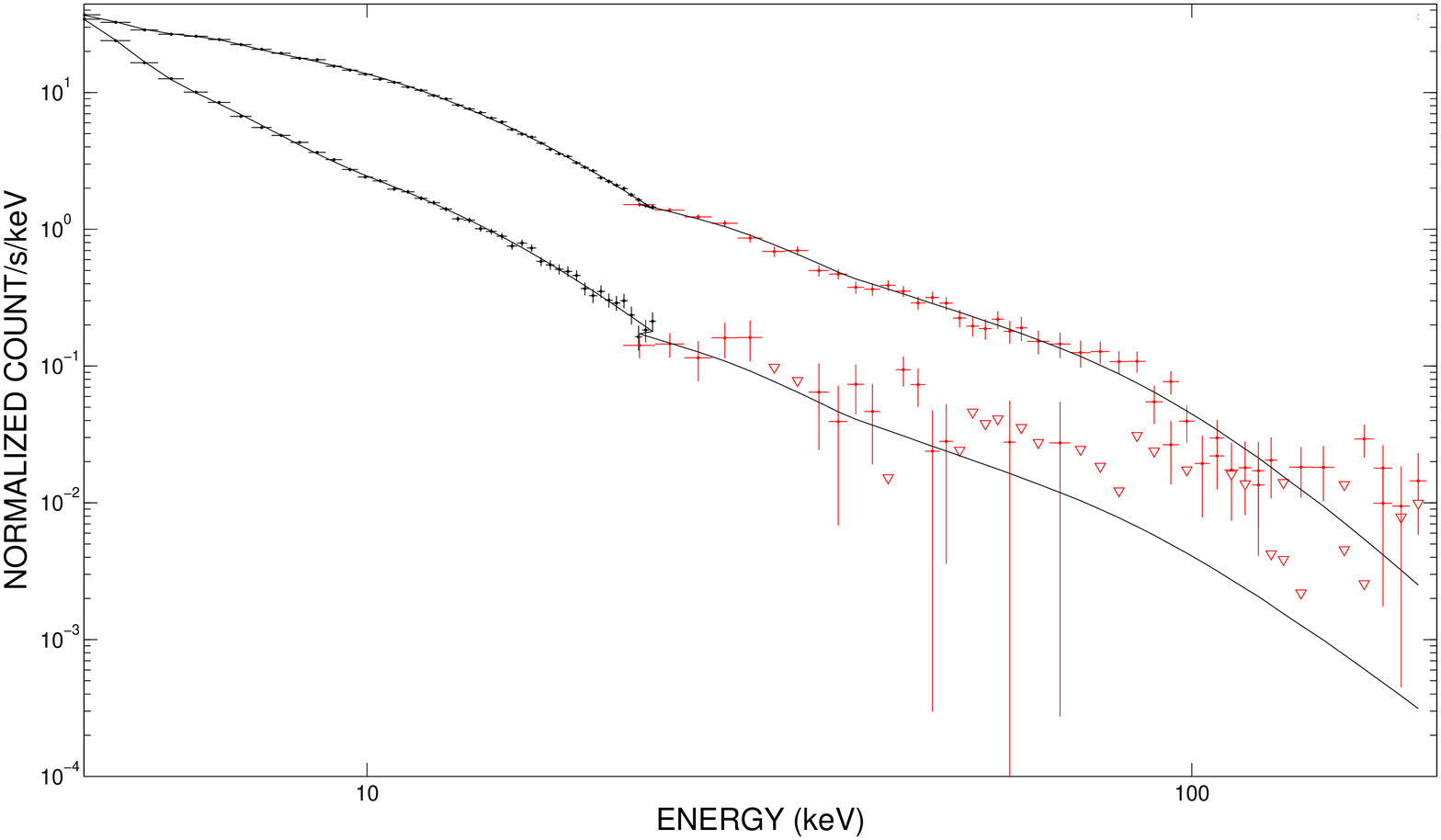} 
\caption{Examples of spectral fits resulted from the combined PCA and HEXTE spectra taken during the 2009 outburst of H1743-322. For both spectra we used a model consisting of interstellar absorption, a Gaussian emission line and a  {\tt pexrav} model. The top spectrum corresponds to the hardest observation of 2009 outburst (Obs. \#1). The lower spectrum corresponds to the softest spectrum of 2009 outburst (Obs. \#26). In this latter spectrum a disk blackbody component had to be added to the model.
} \label{fig:spettri_2009}
\end{figure}

Not all the components of the model are present trough the outburst evolution. At the beginning and at the end of the outburst (Observations \#1 to \#6 and Observations \#35 to \#44), no significant disk emission is observed in the spectra (see Fig. \ref{fig:spettri_2009}, top spectrum). Here, the model used did not contain the disk component and it yielded an average $\chi^{2}_{\mathrm{red}}$ of $\sim$ 1.02 for 81 degrees of freedom (d.o.f.). During the central part of the outburst (from $\#7$ to $\#34$), the disk component is present, yielding an average reduced $\chi ^2 $ of 1.2 for 79 d.o.f. (see Fig. \ref{fig:spettri_2009}, bottom spectrum). 
The time evolution of the spectral parameters is shown in Fig. \ref{fig:parametri} (panels c2 to g2). The observations started in the HIMS. During the first days of the HIMS, the source exhibited a rapid rise in flux, which reached the maximum value of the outburst in less than 8 days. 

From Fig. \ref{fig:parametri} we notice that, as the source is moving through the HIMS, the photon index rises from $\sim$ 1.7 to $\sim$ 2.4 in less than 4 days (from Obs. \#1 to Obs. \#6). This progressive softening of the source is possibly associated with the rise of a soft component related to the accretion disk emission. Despite the softening, the multicolor disk component is not directly observable in the spectrum until observation $\#7$, probably because of the low temperature of the accretion disk. From observation $\#7$ the photon index remains between $\sim$ 2.3 and $\sim$ 3.1. After Obs $\#35$, when the source is about to come back to the LHS, the photon index sensibly decreases to values lower than $\sim$ 2.1. As it is expected, the higher values of the photon index (over $\sim $ 2.1) correspond to the phases in which the multicolor disk component is visible in the spectra (from Obs. \#7 to Obs. \#34). 
The high-energy cut-off is present during the HIMS, from observation $\#1$ to $\#5$. During this period it ranges between $\sim$ 140 keV and $\sim$ 200 keV. The large error bars no not allow to check for variability in the cutoff energy. Because of the response of HEXTE at its highest energies, the cutoff values are not well constrained over $\sim$180 keV. From observation $\#5$ on the cutoff energy, if present, is over 200 keV. In order to check for the absence/presence of an high-energy cut-off in proximity of the HIMS/SIMS transition, we averaged spectra with similar hardness and compatible spectral parameters to accumulate better statistics. The mean spectra show values of the high-energy cutoff consistent with the values from single observations, although slightly better constrained. Although the large error bars still make it difficult to pinpoint a precise evolution of the high-energy cutoff, it is possible to say that it seems to decrease and then increase as the source approaches the soft states. The results are summarized in table \ref{tab:somme} and the mean values obtained for the cutoff are shown in Fig. \ref{fig:parametri} (panel f2).

The disk component is not present between observations \#1 to \#6. When it appears, in correspondence to the transition to the SIMS (Obs. $\#7$), the inner disk radius appears to be consistent with constant \footnote{The radius varies between $\sim$ 25 and $\sim$80 km. Assuming a distance of $\sim$ 10 kpc and a $\sim$ 65$^o$ disk inclination angle (\citealt{MacC2009})}). The values for Obs. $\#28$ to $\#33$ are not well constrained because of the low photon count. During this phase, the source goes back to harder states and the disk emission is expected to drop and disappear again from the PCA band. The temperature of the disk decreases as the source softes, moving from $\sim$ 0.9 keV to $\sim$ 0.5 keV. 
The reflection scaling factor remains almost constant during the whole outburst, showing slightly higher values when the disk component is visible. 
This is expected if the reflected emission is due to reprocessing of the corona emission by the disc.
%----------------------------------------------------------------------------------
%----------------------------------------------------------------------------------

\subsection{The 2008 outburst of H1743-322}
We extended the analysis described for the 2009 outburst to the 2008 outburst. 

As can be seen in Fig. \ref{fig:HID} (red track), the source traces a different path during the 2008 outburst. % The results of the {\it RXTE}, {\it Swift} and INTEGRAL data analysis presented by Capitanio et al. (2009) show that during the outbursts phase only the HIS was clearly observed by RXTE.
Following the HID (Fig. \ref{fig:HID}, red track), we see that the source moves horizontally to the left, then jumps to a softer state, from which it slowly returns to the hard track along a diagonal path. As one can see, the softest points of the second outburst reach intermediate values of the hardness ($\sim$ 0.5), corresponding to the HIMS of the 2009 outburst. This is clearly seen in Fig. \ref{fig:parametri} (panel b1 and b2): the 2008 outburst (red track) is harder than the 2009 one. 
%%_____________________BEGIN________TABLE 4____________________________%%
\begin{table*} 
\renewcommand{\arraystretch}{1.3} 
\begin{center} 
\begin{tabular}{|c|c|c|c|c|c|c|c|} 
\multicolumn{8}{|c|}{2008 Outburst} \\
\hline 
Obs no. & obs ID & MJD & color & PCU2 Counts s$^{-1}$ & rms & QPO & State \\
\hline \hline 
1 & 93427-01-09-00 & 54742.98 & $ 0.878 \pm 0.006 $ & $ 200.6 \pm 0.6 $ & $ 29.7 \pm 0.8 $ & C & HIMS \\
2 & 93427-01-09-01 & 54746.51 & $ 0.870 \pm 0.004 $ & $ 215.8 \pm 0.4 $ & $ 30.4 \pm 0.3 $ & C & HIMS \\
3 & 93427-01-09-03 & 54747.49 & $ 0.860 \pm 0.004 $ & $ 214.6 \pm 0.4 $ & $ 29.7 \pm 0.3 $ & C & HIMS \\
4 & 93427-01-09-02 & 54748.21 & $ 0.857 \pm 0.005 $ & $ 220.1 \pm 0.5 $ & $ 29.7 \pm 0.5 $ & C & HIMS \\
5 & 93427-01-10-00 & 54750.37 & $ 0.846 \pm 0.006 $ & $ 219.8 \pm 0.6 $ & $ 29.6 \pm 0.5 $ & C & HIMS \\
6 & 93427-01-10-01 & 54752.26 & $ 0.838 \pm 0.004 $ & $ 221.6 \pm 0.5 $ & $ 29.6 \pm 0.4 $ & C & HIMS \\
7 & 93427-01-10-02 & 54755.23 & $ 0.824 \pm 0.006 $ & $ 209.1 \pm 0.6 $ & $ 29.6 \pm 0.5 $ & C & HIMS \\
8 & 93427-01-11-00 & 54756.25 & $ 0.818 \pm 0.004 $ & $ 208.1 \pm 0.4 $ & $ 28.9 \pm 0.2 $ & C & HIMS \\
9 & 93427-01-11-01 & 54758.15 & $ 0.800 \pm 0.004 $ & $ 201.3 \pm 0.4 $ & $ 28.0 \pm 0.2 $ & C & HIMS \\
10 & 93427-01-11-03 & 54762.14 & $ 0.526 \pm 0.002 $ & $ 249.8 \pm 0.5 $ & $ 16.3 \pm 0.4 $ & C & HIMS \\
11 & 93427-01-12-00 & 54764.24 & $ 0.478 \pm 0.003 $ & $ 260.0 \pm 0.6 $ & $ 9.8 \pm 1.0 $ & - & HIMS \\
12 & 93427-01-12-03 & 54765.49 & $ 0.491 \pm 0.003 $ & $ 235.8 \pm 0.5 $ & $ 12.2 \pm 0.4 $ & C & HIMS \\
13 & 93427-01-12-01 & 54766.09 & $ 0.487 \pm 0.003 $ & $ 233.0 \pm 0.6 $ & $ 12.1 \pm 0.5 $ & C & HIMS \\
14 & 93427-01-12-04 & 54767.84 & $ 0.608 \pm 0.003 $ & $ 176.9 \pm 0.4 $ & $ 19.5 \pm 0.3 $ & C & HIMS \\
15 & 93427-01-12-02 & 54768.64 & $ 0.647 \pm 0.004 $ & $ 165.9 \pm 0.5 $ & $ 21.9 \pm 0.8 $ & C & HIMS \\
16 & 93427-01-12-05 & 54769.14 & $ 0.668 \pm 0.005 $ & $ 159.3 \pm 0.5 $ & $ 21.6 \pm 0.5 $ & C & HIMS \\
17 & 93427-01-13-00 & 54770.12 & $ 0.695 \pm 0.003 $ & $ 148.5 \pm 0.3 $ & $ 23.4 \pm 0.3 $ & C & HIMS \\
18 & 93427-01-13-05 & 54770.40 & $ 0.709 \pm 0.004 $ & $ 143.8 \pm 0.3 $ & $ 24.0 \pm 0.3 $ & C & HIMS \\
19 & 93427-01-13-04 & 54771.76 & $ 0.719 \pm 0.004 $ & $ 134.4 \pm 0.3 $ & $ 23.7 \pm 0.3 $ & C & HIMS \\
20 & 93427-01-13-01 & 54772.15 & $ 0.727 \pm 0.006 $ & $ 131.1 \pm 0.5 $ & $ 24.3 \pm 0.6 $ & C & HIMS \\
21 & 93427-01-13-02 & 54773.27 & $ 0.728 \pm 0.005 $ & $ 125.8 \pm 0.4 $ & $ 23.8 \pm 0.5 $ & C & HIMS \\
22 & 93427-01-13-06 & 54774.64 & $ 0.732 \pm 0.004 $ & $ 118.7 \pm 0.3 $ & $ 23.9 \pm 0.3 $ & C & HIMS \\
23 & 93427-01-13-03 & 54775.56 & $ 0.760 \pm 0.005 $ & $ 114.1 \pm 0.3 $ & $ 24.4 \pm 0.4 $ & C & HIMS \\
24 & 93427-01-14-00 & 54777.86 & $ 0.750 \pm 0.005 $ & $ 111.6 \pm 0.3 $ & $ 23.7 \pm 0.4 $ & C & HIMS \\
25 & 93427-01-14-01 & 54778.77 & $ 0.765 \pm 0.006 $ & $ 110.4 \pm 0.4 $ & $ 22.5 \pm 1.2 $ & C & HIMS \\
26 & 93427-01-14-02 & 54779.04 & $ 0.765 \pm 0.006 $ & $ 110.0 \pm 0.4 $ & $ 24.6 \pm 0.4 $ & C & HIMS \\
27 & 93427-01-14-03 & 54780.02 & $ 0.779 \pm 0.006 $ & $ 108.2 \pm 0.4 $ & $ 25.1 \pm 1.0 $ & C & HIMS \\
28 & 93427-01-14-04 & 54781.78 & $ 0.750 \pm 0.006 $ & $ 108.8 \pm 0.4 $ & $ 25.6 \pm 0.6 $ & C & HIMS \\
29 & 93427-01-14-05 & 54782.89 & $ 0.727 \pm 0.005 $ & $ 110.2 \pm 0.4 $ & $ 23.7 \pm 0.4 $ & C & HIMS \\
30 & 93427-01-14-06 & 54783.81 & $ 0.741 \pm 0.006 $ & $ 106.6 \pm 0.4 $ & $ 23.0 \pm 0.4 $ & C & HIMS \\
31 & 93427-01-15-00 & 54784.45 & $ 0.763 \pm 0.006 $ & $ 101.0 \pm 0.3 $ & $ 23.9 \pm 0.6 $ & C & HIMS \\
32 & 93427-01-15-01 & 54785.70 & $ 0.754 \pm 0.006 $ & $ 95.8 \pm 0.4 $ & $ 22.8 \pm 0.5 $ & C & HIMS \\
33 & 93427-01-15-02 & 54786.48 & $ 0.768 \pm 0.005 $ & $ 90.4 \pm 0.3 $ & $ 22.6 \pm 0.5 $ & C & HIMS \\
34 & 93427-01-15-03 & 54787.73 & $ 0.783 \pm 0.007 $ & $ 82.1 \pm 0.3 $ & $ 21.4 \pm 0.6 $ & C & HIMS \\
35 & 93427-01-15-04 & 54788.64 & $ 0.799 \pm 0.008 $ & $ 72.4 \pm 0.3 $ & $ 21.3 \pm 1.1 $ & C & HIMS \\
36 & 93427-01-15-06 & 54789.49 & $ 0.805 \pm 0.008 $ & $ 70.2 \pm 0.3 $ & $ 23.2 \pm 1.0 $ & C & HIMS \\
37 & 93427-01-15-05 & 54788.84 & $ 0.816 \pm 0.008 $ & $ 65.0 \pm 0.3 $ & $ 20.6 \pm 1.0 $ & C & HIMS \\
\hline 
\end{tabular}
\caption{The columns are: observation number for 2008 outburst, RXTE observation ID, MJD, PCU2 count rate, hardness ratio, integrated fractional rms (0.001 - 64 Hz), QPO type, and state according to Belloni (2008).}\label{tab:states_2008} 
\end{center} 
\end{table*}
%%_____________________END________TABLE 4____________________________%%

\subsubsection{Timing analysis}\label{sec:timing_2008}
The timing analysis yielded the following results:
\begin{itemize}
\item Observations from $\#1$ to $\#10$ and from $\#12$ to $\#37$ show a high level of aperiodic variability in the form of strong band-limited noise components (flat top noise shape). The total integrated fractional rms is in the range $16 - 30\%$. The PDS can be decomposed in a number of Lorentzian components, one of which takes the form of a type-C QPO peak.
These observations correspond to the triangles in Fig. \ref{fig:HID}. 
\item Observation $\#11$ corresponds to a slightly weaker variability ($\sim 10\%$ fractional rms). The PDS is consistent with a zero-centered Laurentian, rhough noisy, with no QPOs. This observation is the only one that does not show any QPO. 
This observation is marked with an empty triangle in the HID of Fig. \ref{fig:HID} (red track). 
\end{itemize}
Results are summarized in Tab \ref{tab:timing_tab_2008}. 
%%_____________________BEGIN________TABLE 5____________________________%%
\begin{table*}
\renewcommand{\arraystretch}{1.3}
\begin{center}
\begin{tabular}{|c|c|c|c|}
\multicolumn{4}{|c|}{2008 Outburst} \\
\hline 
\#obs ID & Noise Type & QPO type & RMS (in \%) \\
\hline
\hline
\#1 to \#10, \#12 to \#37 & strong band limited noise & C & 16-30\% \\
\#11 & weak band limited noise& - & 10\% \\
\hline
\end{tabular}
\caption{Timing properties seen in the PDS of each observation from the 2008 outburst. }\label{tab:timing_tab_2008}
\end{center}
\end{table*}
%%_____________________BEGIN________TABLE 5____________________________%%

On the basis of the PDS properties, all that observations can be classified as HIMS. This fact is supported by the hardness ratio values observed for the corresponding spectra. 
Observation $\#11$ shows timing properties which are consistent with the HIMS, even though no Type-C QPO is observed. However the non-detection is compatible with neighboring observations (see C09 for details).

%%_____________________BEGIN________TABLE 6____________________________%%
\begin{table*} 
\renewcommand{\arraystretch}{1.3} 
\begin{center} 
\begin{tabular}{|c|c|c|c|c|c|} 
\multicolumn{6}{|c|}{2008 Outburst} \\
\hline 
obs no. & kT (keV) & R (Km) & $\Gamma$ & $E_f$ & $\Omega$ \\
\hline 
1 & $ - $ & $ - $ & $ 1.64 _{- 0.04 }^{+ 0.04 } $ & $ > 200 $ & $ 0.6 _{- 0.1 }^{+ 0.2 } $ \\
2 & $ - $ & $ - $ & $ 1.66 _{- 0.03 }^{+ 0.03 } $ & $ > 200 $ & $ 0.7 _{- 0.1 }^{+ 0.1 } $ \\
3 & $ - $ & $ - $ & $ 1.67 _{- 0.03 }^{+ 0.03 } $ & $ > 200 $ & $ 0.6 _{- 0.1 }^{+ 0.1 } $ \\
4 & $ - $ & $ - $ & $ 1.67 _{- 0.03 }^{+ 0.03 } $ & $ > 200 $ & $ 0.6 _{- 0.1 }^{+ 0.1 } $ \\
5 & $ - $ & $ - $ & $ 1.68 _{- 0.04 }^{+ 0.04 } $ & $ 195 _{- 56 }^{+ 115 } $ & $ 0.5 _{- 0.1 }^{+ 0.2 } $ \\
6 & $ - $ & $ - $ & $ 1.68 _{- 0.03 }^{+ 0.03 } $ & $ 183 _{- 43 }^{+ 74 } $ & $ 0.4 _{- 0.1 }^{+ 0.1 } $ \\
7 & $ - $ & $ - $ & $ 1.76 _{- 0.04 }^{+ 0.04 } $ & $ > 200 $ & $ 0.6 _{- 0.2 }^{+ 0.2 } $ \\
8 & $ - $ & $ - $ & $ 1.74 _{- 0.03 }^{+ 0.03 } $ & $ > 200 $ & $ 0.5 _{- 0.1 }^{+ 0.1 } $ \\
9 & $ - $ & $ - $ & $ 1.79 _{- 0.03 }^{+ 0.03 } $ & $ > 200 $ & $ 0.6 _{- 0.1 }^{+ 0.1 } $ \\
10 & $ - $ & $ - $ & $ 2.58 _{- 0.03 }^{+ 0.03 } $ & $ > 200 $ & $ 1.4 _{- 0.2 }^{+ 0.2 } $ \\
11 & $ 0.74 _{- 0.07 }^{+ 0.13 } $ & $ 36 _{- 12 }^{+ 52 } $ & $ 2.22 _{- 0.11 }^{+ 0.17 } $ & $ > 200 $ & $ 0.2 _{- 0.2 }^{+ 0.4 } $ \\
12 & $ 0.74 _{- 0.12 }^{+ 0.09 } $ & $ 30 _{- 10 }^{+ 113 } $ & $ 2.27 _{- 0.17 }^{+ 0.19 } $ & $ > 200 $ & $ 0.4 _{- 0.4 }^{+ 0.6 } $ \\
13 & $ 0.73 _{- 0.19 }^{+ 0.21 } $ & $ 34 _{- 12 }^{+ 124 } $ & $ 2.20 _{- 0.29 }^{+ 0.32 } $ & $ > 200 $ & $ 0.3 _{- 0.3 }^{+ 0.9 } $ \\
14 & $ 0.45 _{- 0.06 }^{+ 0.19 } $ & $ >77 $ & $ 2.11 _{- 0.23 }^{+ 0.09 } $ & $ > 200 $ & $ 0.4 _{- 0.2 }^{+ 0.3 } $ \\
15 & $ - $ & $ - $ & $ 2.22 _{- 0.04 }^{+ 0.04 } $ & $ > 200 $ & $ 0.9 _{- 0.3 }^{+ 0.2 } $ \\
16 & $ - $ & $ - $ & $ 2.15 _{- 0.05 }^{+ 0.05 } $ & $ > 200 $ & $ 0.8 _{- 0.3 }^{+ 0.3 } $ \\
17 & $ - $ & $ - $ & $ 2.08 _{- 0.03 }^{+ 0.03 } $ & $ > 200 $ & $ 0.7 _{- 0.2 }^{+ 0.1 } $ \\
18 & $ - $ & $ - $ & $ 2.06 _{- 0.02 }^{+ 0.03 } $ & $ > 200 $ & $ 0.8 _{- 0.2 }^{+ 0.2 } $ \\
19 & $ - $ & $ - $ & $ 1.99 _{- 0.04 }^{+ 0.03 } $ & $ > 200 $ & $ 0.7 _{- 0.2 }^{+ 0.1 } $ \\
20 & $ - $ & $ - $ & $ 1.97 _{- 0.05 }^{+ 0.05 } $ & $ > 200 $ & $ 0.4 _{- 0.2 }^{+ 0.2 } $ \\
21 & $ - $ & $ - $ & $ 2.01 _{- 0.04 }^{+ 0.04 } $ & $ > 200 $ & $ 0.7 _{- 0.1 }^{+ 0.2 } $ \\
22 & $ - $ & $ - $ & $ 1.98 _{- 0.03 }^{+ 0.03 } $ & $ > 200 $ & $ 0.6 _{- 0.2 }^{+ 0.1 } $ \\
23 & $ - $ & $ - $ & $ 1.89 _{- 0.04 }^{+ 0.03 } $ & $ > 200 $ & $ 0.3 _{- 0.1 }^{+ 0.1 } $ \\
24 & $ - $ & $ - $ & $ 1.90 _{- 0.04 }^{+ 0.02 } $ & $ > 200 $ & $ 0.4 _{- 0.2 }^{+ 0.2 } $ \\
25 & $ - $ & $ - $ & $ 1.91 _{- 0.06 }^{+ 0.05 } $ & $ > 200 $ & $ 0.6 _{- 0.3 }^{+ 0.2 } $ \\
26 & $ - $ & $ - $ & $ 1.88 _{- 0.04 }^{+ 0.04 } $ & $ > 200 $ & $ 0.5 _{- 0.2 }^{+ 0.2 } $ \\
27 & $ - $ & $ - $ & $ 1.85 _{- 0.04 }^{+ 0.04 } $ & $ > 200 $ & $ 0.4 _{- 0.2 }^{+ 0.2 } $ \\
28 & $ - $ & $ - $ & $ 1.92 _{- 0.02 }^{+ 0.04 } $ & $ > 200 $ & $ 0.5 _{- 0.2 }^{+ 0.2 } $ \\
29 & $ - $ & $ - $ & $ 1.97 _{- 0.05 }^{+ 0.05 } $ & $ > 200 $ & $ 0.5 _{- 0.2 }^{+ 0.3 } $ \\
30 & $ - $ & $ - $ & $ 1.95 _{- 0.06 }^{+ 0.05 } $ & $ > 200 $ & $ 0.7 _{- 0.3 }^{+ 0.3 } $ \\
31 & $ - $ & $ - $ & $ 1.90 _{- 0.04 }^{+ 0.04 } $ & $ > 200 $ & $ 0.5 _{- 0.2 }^{+ 0.2 } $ \\
32 & $ - $ & $ - $ & $ 1.92 _{- 0.06 }^{+ 0.06 } $ & $ > 200 $ & $ 0.5 _{- 0.2 }^{+ 0.3 } $ \\
33 & $ - $ & $ - $ & $ 1.89 _{- 0.04 }^{+ 0.04 } $ & $ > 200 $ & $ 0.5 _{- 0.1 }^{+ 0.3 } $ \\
34 & $ - $ & $ - $ & $ 1.81 _{- 0.05 }^{+ 0.05 } $ & $ > 200 $ & $ 0.3 _{- 0.2 }^{+ 0.3 } $ \\
35 & $ - $ & $ - $ & $ 1.80 _{- 0.06 }^{+ 0.06 } $ & $ > 200 $ & $ 0.5 _{- 0.2 }^{+ 0.3 } $ \\
36 & $ - $ & $ - $ & $ 1.80 _{- 0.06 }^{+ 0.05 } $ & $ > 200 $ & $ 0.5 _{- 0.2 }^{+ 0.3 } $ \\
37 & $ - $ & $ - $ & $ 1.82 _{- 0.05 }^{+ 0.06 } $ & $ > 200 $ & $ 0.6 _{- 0.2 }^{+ 0.3 } $ \\
\hline
\end{tabular}
\caption{Spectral parameters of the 2008 outburst of H1743-322. Columns are: observation number, inner disc temperature kT, inner disc radius R (assuming a distance of 10 kpc and an inclination of 65 $^o$), photon index $\Gamma$, fold Energy E$_{fold}$ (corresponding to high energy cutoff), reflection factor $\Omega$. The inner disk radii are calculated from the disk-blackbody normalization, defined as $(\frac{R_{in}/ km}{D/10 kpc})^2 cos \Theta$, where R$_{in}$ is the inner disc radius (km), D is the distance to the source (kpc) and $\Theta$ is the inclination angle of the disk.}\label{tab:spettrali_2008} 
\end{center} 
\end{table*}
%%_____________________BEGIN________TABLE 6____________________________%%

\subsubsection{Spectral analysis}\label{sec:spectral_analysis_2008}
We applied to the 2008 outburst the same approach we adopted for the 2009 outburst. A model consisting in a simple cutoff powerlaw plus disk-blackbody component as well as a broken powerlaw with high-energy cutoff plus disk-blackbody was firstly tested. As for the 2009 data, the resulting fits are good for the central part of the outburst, but they are not for the beginning and ending phases (see Sec. \ref{sec:spectral_analysis}). Again, the resulting normalization of the disk-blackbody component yielded inner radius values too small to be accepted.
%With the same assumption described in Sec. \ref{sec:spectral_analysis}, we found that the resulting normalizations ranged between $\sim$ 60 and $\sim$ 100, which yielded inner disk radii ranging between $\sim$ 9 and $\sim$ 12 km. This means that the maximum inner disk radius found equals 0.42 gravitational radii, which is a non acceptable value from a physical point of view.

\begin{figure}
\includegraphics[width=8.5cm]{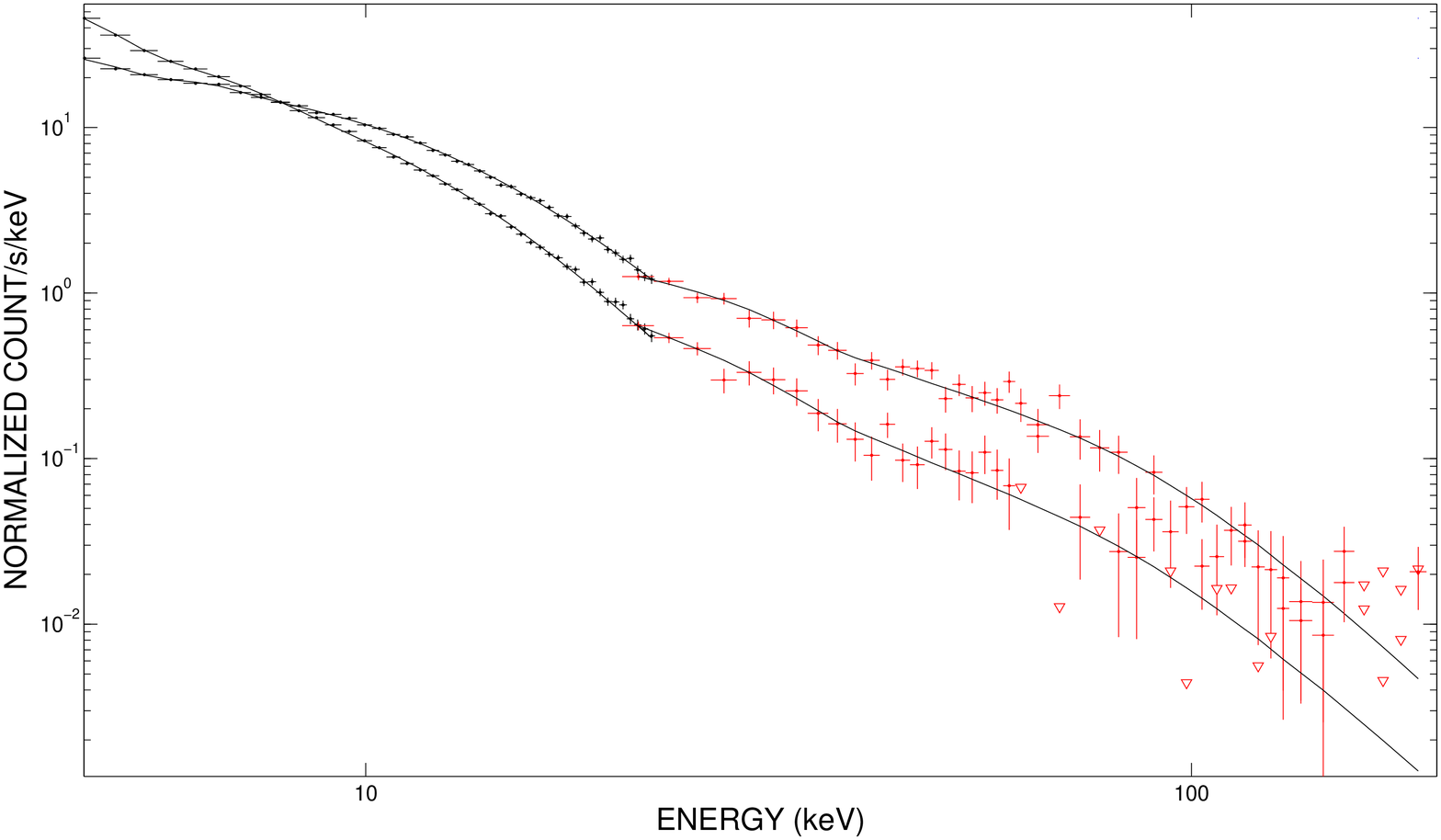}
\caption{Examples of spectral fits from the combined PCA and HEXTE spectra taken during the 2008 outburst of H1743-322. For both spectra we used a model consisting of interstellar absorption, a Gaussian emission line and a {\tt pexrav} model. The top spectrum corresponds to the hardest observation of 2008 outburst (i.e. Obs. \#1). The lower spectrum corresponds to the softest spectrum of 2009 outburst (i.e. Obs. \#11). In this latter spectrum a disk blackbody component had to be added to the fit.} \label{fig:spettri_2008}
\end{figure}

We therefore adopted the same model as for the 2009 outburst  (pexrav+disk-blackbody) .
%For the 2008 outburst the addition of the absorption edge component gave better results than for the 2009 outburst. The resulting optical depth values were similar to those found for the 2009 data, but the edge energies resulted better constrained. Even so, the absorption edge component not clearly significant and therefore was not included. 
We found that the disk black-body component is required only for the very central part of the outburst in correspondence with the softest points of the HID (Obs. \#11 to \#14). This is probably due to the faintness of the accretion disk, which is likely too cold to be clearly detected by PCA. The fits for those observations yield an average $\chi^{2}_{\mathrm{red}}$ of 1.08 for 79 d.o.f. (see Fig. \ref{fig:spettri_2008}, bottom spectrum). In observations \#11 to \#14 the disk blackbody component has to be added to the model to account of the soft excess observed.
All the other observations could be fitted without a disk blackbody component, yielding an average $\chi^{2}_{\mathrm{red}}$ of 0.98 for 81 d.o.f. (see Fig. \ref{fig:spettri_2008}, top spectrum). 
%In addition, the inner disk radius values that Capitanio et al. report are not consistent with ours. Their fits give lower normalizations associated to higher temperatures measured at the inner disk radius. Their solution should not be ruled out \textit{a priori}, given that different scenarios are possible. Nevertheless the values that they obtain are not acceptable since the inner disk radii calculated from the normalization of the disk black-body component are too small to be physically acceptable. We attribute the difference between our results and the results of Capitanio at al. to the oversimplification of the spectral model they used. Capitanio et al. (2009) do not consider a reflection component, which we find necessary to obtain good fits. As we described before, we tried to use the model described by Capitanio et al. (2009), but it was not enough to describe our spectra.

The spectral evolution of H1743-322 during its 2008 outburst is shown in Fig.  \ref{fig:parametri} (red tracks) and the corresponding values are reported in Table. \ref{tab:spettrali_2008}. 
As expected during the HIMS, the photon index remains nearly constant (ranging between 1.6 and 1.7). After that, it jumps to $\sim$ 2.6 in less then two days (at Obs. $\#10$). This is due to the appearance of a soft spectral component probably associated to the accretion disk, which becomes directly observable from observation $\#11$. After observation $\#10$ the photon index moves back to slightly lower values ($\sim$ 2.2) and then moves down to $\sim$ 1.8. This is the typical behaviour observed during the HIMS, when the system is backing towards the LHS.
The high-energy cutoff is not well constrained for the 2008 outburst. The fits give values which in all the cases but two (observation $\#5$ and $\#6$) are too high to be considered acceptable (see Tab. \ref{tab:spettrali_2008}). As for the 2009 outburst, we averaged spectra corresponding to similar hardness ratio in order to acquire better statistics. The mean spectra obtained show values of the high-energy cutoff consistent with the values of the single observation spectra and still too high to be considered reliable. However the mean spectrum of observations \#5 and \#6 shows a clearer cutoff at $\sim$ 180 keV. The results are summarized in table \ref{tab:somme} and the mean values obtained for the cutoff are shown in panel f1 in Fig. \ref{fig:parametri}).
 
The reflection-scaling factor remains nearly constant along the outburst. This could be explained by a lack of disk flux. Given that the reflection component is directly tied to the disk emission (see Sec. \ref{sec:discussion_2008}), the low reflection component is consistent to the presence of a  faint disk (i.e. low disk flux, see Fig. \ref{fig:components}).

%----------------------------------------------------------------------------------
%----------------------------------------------------------------------------------
%%_____________________BEGIN________TABLE 7____________________________%%
\begin{table*} 
\renewcommand{\arraystretch}{1.3} 
\begin{center} 
\begin{tabular}{|c|c|c|c|c|c|} 
\multicolumn{6}{|c|}{2008 Outburst} \\
\hline 
obs no. & kT (keV) & R (Km) & $\Gamma$ & $E_f$ & $\Omega$ \\
\hline 
$ \#1,\#2 $ & $ - $ & $ - $ & $ 1.65 _{- 0.03 }^{+ 0.03 } $ & $ >200$ & $ 0.61 _{- 0.09 }^{+ 0.09 } $ \\
$ \#3,\#4 $ & $ - $ & $ - $ & $ 1.67 _{- 0.02 }^{+ 0.02 } $ & $ >200$ & $ 0.60 _{- 0.07 }^{+ 0.07 } $ \\
$ \#5,\#6 $ & $ - $ & $ - $ & $ 1.68 _{- 0.03 }^{+ 0.03 } $ & $ 190 _{- 38 }^{+ 57 } $ & $ 0.44 _{- 0.09 }^{+ 0.09 } $ \\
$ \#7,\#8 $ & $ - $ & $ - $ & $ 1.75 _{- 0.03 }^{+ 0.03 } $ & $ >200$ & $ 0.6 _{- 0.1 }^{+ 0.1 } $ \\
\hline 
\multicolumn{6}{|c|}{} \\ 
\multicolumn{6}{|c|}{2009 Outburst} \\ 
\hline 
$ \#1,\#2 $ & $ - $ & $ - $ & $ 1.73 _{- 0.02 }^{+ 0.02 } $ & $ 166 _{- 23 }^{+ 30 } $ & $ 0.60 _{- 0.07 }^{+ 0.07 } $ \\
$ \#3,\#4 $ & $ - $ & $ - $ & $ 1.77 _{- 0.02 }^{+ 0.02 } $ & $ 153 _{- 23 }^{+ 32 } $ & $ 0.53 _{- 0.09 }^{+ 0.09 } $ \\
$ \#4,\#5 $ & $ - $ & $ - $ & $ 1.88 _{- 0.02 }^{+ 0.02 } $ & $ 196 _{- 34 }^{+ 50 } $ & $ 0.52 _{- 0.07 }^{+ 0.07 } $ \\
\hline
\end{tabular}
\caption{Spectral parameters coming from spectra averaged across multiple observations (see text). Columns are: observation number, inner disc temperature kT, inner disc radius R (assuming a distance of 10 kpc and an inclination of 65$^o$), photon index $\Gamma$, fold Energy E$_{fold}$ (corresponding to high-energy cutoff), reflection factor $\Omega$. The inner disk radii are calculated from the disk-blackbody normalization, defined as $(\frac{R_{in}/ km}{D/10 kpc})^2 cos \Theta$, where R$_{in}$ is the inner disc radius (km), D is the distance to the source (kpc) and $\Theta$ is the inclination angle of the disk.}\label{tab:somme} 
\end{center} 
\end{table*} 
%%_____________________END__________TABLE_7____________________________%%

\section{Discussion}\label{sec:discussion}
The different behaviour shown by H1743-322 during its outburst evolution in 2008 and 2009 allow us to study the evolution of the spectral parameters of the system at different accretion regimes.

\subsection{The 2009 outburst}
Our results  show that H1743-322 underwent a typical outburst between 2009 May 29 and 2009 July 27. The source probably went through a (missed) initial LHS and then crossed the HID following the upper horizontal branch (HIMS) of the HID. After a very short SIMS and a relatively long permanence in the HSS, the source went back to the LHS passing through the lower horizontal branch (HIMS). 
During the softening of the source two main transitions, a secondary transition and a backward transition have been identified. All the transitions took place at hardness values consistent with what is observed in other sources.
To date now, the rising phase has not been observed for this source in any of the outbursts covered by RXTE. This is probably due to the fact that the transition from quiescence to HIMS is very fast and therefore difficult to be observed.

The spectral parameters evolved consistently with the ones previously observed  in the source  (see \citealt{Prat2009}) and other sources (see e.g. \citealt{Motta2009}, \citealt{DelSanto2009}, \citealt{Belloni2005a} for GX 339-4; see \citealt{Debnath2009} for GRO J1655-40): 
\begin{itemize}
\item the photon index showed the expected evolution in relation to the variation of the spectral components. The appearance of the soft disk-blackbody component, together with the progressive cooling of the Comptonizing medium in the spectrum, is probably the reason for the photon index rising. Independently of whether the inner radius of the accretion disk moves inward or not, more soft photons will be emitted by the disk as the source becomes brighter. The photon input to the Comptonizing medium will therefore increase. This will steepen the spectrum and will cool the population of electrons (see \citealt{Sunyaev1980}). As observed for GX 339-4, changes in the photon index are observed across the transition from the HIMS to the HSS. It rises during the HIMS and becomes constant when the source reaches the HSS. Then it stays constant until the source enters again the HIMS, coming back to the LHS. As the source begins the hardening, the photon index decreases again.
\item The high-energy cutoff evolution could not be observed in detail, mainly because of the lack of observation in the LHS (where the high-energy cutoff is expected to be observable (e.g. \citealt{Motta2009}). However, the values we obtained from our fits are consistent with the presence of a high-energy cutoff before the transition towards the soft states. 
\item The inner disk radius is consistent with remaining constant and small (being around 50 Km), although, given the large error bars, a scenario in which the inner disk radius moves inward or outward cannot be ruled out. Observation made at lower energies and during earlier phases of the outburst evolution could provide further information to define properly the geometry of the system.
\item As it is expected, the reflection scaling factor is correlated to the photon index and to the disk parameters.

The strength of the reflected component depends on the fraction of the Comptonized radiation intercepted by the accretion disk. The latter is defined by the geometry of the accretion flow, i.e. by the solid angle $\Omega_{disk}$ subtended by the accretion disk as seen from the Comptonizing region, and by the ionization state of the disk.
%. In addition, the spectrum of the reflected emission depends on the ionization state of the disk. In particular its low energy part is formed by both the interplay between Thomson scattering and photo-absorption and fluorescence by metals. The problem is further complicated by the fact that the ionization state of the disk can be modified by the Comptonized radiation.
%The observations show that there is a clear correlation between the photon index of the Comptonized radiation $\Gamma$ (i.e. the Comptonization parameter) and the relative amplitude of the reflected component (Gilfanov 2009 and reference therein). 
The bigger is the fraction of Comptonized radiation, the stronger is expected to be the reflection component. As the 
emission that is  Comptonized is produced in the accretion disk, softer spectra (i.e. where the disk emission dominates) have a stronger reflected component, yelding a larger equivalent width of the iron fluorescent line. %The existence of this correlation suggests that there is a positive correlation between the fraction of the Comptonized radiation intercepted by the accretion disk and the energy flux of the soft seed photons to the Comptonization region. %This is a strong argument in favor of the accretion disk being the primary source of soft seed photons to the Comptonization region.
The large amount of reflected radiation is consistent with the relative large observed iron line width (ranging between $\sim$ 0.4 and 1.2 keV). 
\end{itemize}
%%
%% TMB ==========================================================
%
\subsection{The peculiar nature of the 2008 outburst}\label{sec:discussion_2008}

The two observed outburst spanned slightly different luminosity ranges. The maximum luminosity (2.73-200 keV) in the 2008 was 2.45 $\times$ $10^{37}$ erg $s^{-1}$ cm$^{-2}$ while in the 2009 was 2.88 $\times$ $10^{37}$ erg $s^{-1}$ cm$^{-2}$. As it was observed in other sources (see e.g. \citealt{Belloni2010}), the maximum luminosities are correlated with the difference in count rate between the two horizontal branches of the HID, i.e. the higher is the luminosity reached, the bigger is the difference in count rate (\citealt{Maccarone2003}). 

In the 2008 case the maximum luminosity is observed at the very beginning of the outburst, where the hard emission dominates, while in the 2009 case the maximum luminosity is observed in the soft state, where the disk component starts to dominate the spectrum. As most likely, the luminosity is directly related to the accretion rate. The fact that similar luminosities are reached during the two different outbursts is relevant. This could be interpreted as the consequence of two different mechanisms that lead the source towards the maximum luminosity. The difference between the two mechanism might also explain what causes the transition occourrence.

From the analysis shown above, it is clear that the 2008 outburst of H1743-322 was peculiar. 
The HID shape is similar to what is usually observed in other BHTs, but after the usual path through the HIMS, it returned to harder states with no sampling of SIMS/HSS. %This is confirmed by the timing and spectral analysis.
The softest point observed in the HID roughly corresponded to the last part of the HIMS. %The hardness is only a rough indication of the spectral shape and it has been noticed that similar hardness values could correspond to different states even for different outbursts of the same source (\citealt{Belloni2010}). 
%The main difference between the 2009 and the 2008 outbursts is that the latter did not show the typical HID evolution. Only two of the four canonical states are observed.

The timing analysis has pointed out that some changes take place when the source is approaching the softest points in the HID. The fractional rms and the PDS evolve following the behavior observed in several other sources: the rms decreases while the frequencies of the Lorentzian components in the PDS moves towards higher values. No transition to the SIMS \footnote{The transition from HIMS to the SIMS is marked by the appearance of type-B and/or type-A QPO.} is observed before the softest point of the HID is reached (Obs. \#11). 
The softest observation (Obs. \#11) presents the typical rms and hardness values of the HIMS and the PDS display the expected shape (i.e. flat top noise), although slightly noisy.  It might be possible that Obs. \#11 lays close to the transition to the SIMS. As the transition to the soft state is expected to be very fast (see Nespoli et al. 2003) and given that the SIMS is crossed in less than one day during the 2009 outburst, it is possible that the source reached the SIMS and perhaps the HSS during the period where it was not observed by RXTE ($\sim$ 30 hours). 

%belongs to the SIMS, even though no QPOs are detected. One must take into account that Type-A QPOs are usually fainter and broader than the other QPOs, and for this reason it happens that they become detectable only averaging several observations.

The evolution of the spectral parameter during the two different outbursts puts in evidence the lack of observed soft states.
\begin{itemize}
\item As it happened for the 2009 outburst, the photon index slightly increases during the HIMS and then clearly changes its trend once the source reaches the softest point. After that, it did not remain constant as for the 2009 outburst, but suddenly decreased. The track followed by the photon index during the final part of the 2008 outburst is very similar to the one of the 2009 outburst. %The fact that the accretion disk component is never dominant in the spectrum during the outburst is another indication of the lack of the soft states. 
\item As it happened for the 2009 outburst, the high-energy cutoff could not be observed in detail, mainly because of the lack of observation in the LHS (where it is usually lower). However, the values we obtained from our fits are consistent with the presence of a high-energy cutoff before the softest point is reached. The high energy cutoff seems to be slightly higher on average during the 2008 outburst. However, the values are still consistent during the 2008 and 2009 outburst. As we pointed out in Sec. \ref{sec:spectral_analysis}, a trustable value of the high-energy cutoff has been obtained only for a few HIMS observations. For all the other observations it is not present or it is too high to be detected. 
The high-energy cutoff is thought to be related to the temperature of the thermal Comptonizing electrons located in an optically thin corona. According to this, higher cutoff values during the 2008 would suggest that the coronal temperature was higher than for the 2009. Thus, the higher temperature of the corona could have affected the subsequent evolution of the source. % together with the low accretion rate, producing the peculiar 2008 outburst. 
%\item The energy spectrum of the softest observation presents an accretion disk component (which is not present in the previous observations), consistently with what is generally observed during the softest part of the HIMS. As it is expected, the softest observation presents the highest disk flux of the whole outburst, due to the appearance of the disk component. The evolution of the inner disk radius could not be followed in detail, because of the faint emission coming from the disk,  visible for a very short time. The values yielded by the fits are consistent to the values obtained for the 2009 outburst. During the 2009 the disk emission clearly dominates the spectrum during the HSS (for $\sim$ 21 days), yielding a clear evolution for both the temperature and the inner disk radius (see Fig. \ref{fig:components}). During the 2008 outburst the disk never becomes dominant and results observable only in a short period, just in correspondence of the softest part of the HID (for $\sim$ 4 days).
The inner disk radius values we find are not consistent with the ones reported by C09. Their fits give lower normalizations associated to higher temperatures measured at the inner disk radius. The values that they obtain are not acceptable since the inner disk radii calculated from the normalization of the disk black-body component are too small to be physically acceptable. We attribute this difference to the oversimplification of the spectral model they used. C09 do not consider a reflection component, which we find necessary to obtain good fits. As we described before, we tried to use the model described by C09, but it was not enough to describe our spectra.
\end{itemize}

The spectral parameters evolutions at the beginning of the two outbursts are consistent and do not allow to predict the following evolution of the source. The 2008 outburst of H 1743-322, showing only LHS and HIMS, takes place at low luminosity and the lack of soft-state transitions is probably connected to a premature decrease of the mass accretion rate. Although the accretion rate is a likely important parameter to be considered in a transition scenario, another parameter seems to play a role. This parameter, whose nature is still not clear (\citealt{Esin1997}, \citealt{Homan2001}), could drive the source from the LHS to the HIMS and eventually to the HSS. From our study two possibilities stand: either the parameter that drives the transition is not tied to any spectral parameter or during 2008 the transition took place on a short time-scale (less then one day) and therefore RXTE missed it. For the last case to work the system must have stayed in the soft state less than 30 hours. Such a short permanence in the HSS has not been observed in other sources so far and should therefore be explained. However, the peculiarity of the softest 2008 observation could be relevant. As it is observed in other sources (e.g. \citealt{Belloni2005a}), just before the entering into the soft state, the Type-C QPOs typical of  the HIMS observations shows a changing in its shape (i.e. the type-C QPO becomes blunt, see \citealt{Belloni2005}) that could be seen as a loosing of coherence. This behavior is always seen before the appearance of a Type-B QPO, that marks the beginning of the SIMS. What is observed during the 2008 could indicate that the source was at least very close to the transition, even though we cannot state wether the transition took place.

Other cases previously presented in literature as failed outbursts in transient X-ray binaries, are LHS-only outbursts without any sign of state transitions at all (see e.g. \citealt{Brocksopp2004} and \citealt{Sturner2005}). Conversely, the data presented here show that the full pattern (LHS, HIMS, SIMS, HSS) and LHS-only pattern are not the only two possibilities for the temporal evolution of a BHC outburst.

%----------------------------------------------------------------------------------
\begin{figure*}
\includegraphics[width=17cm]{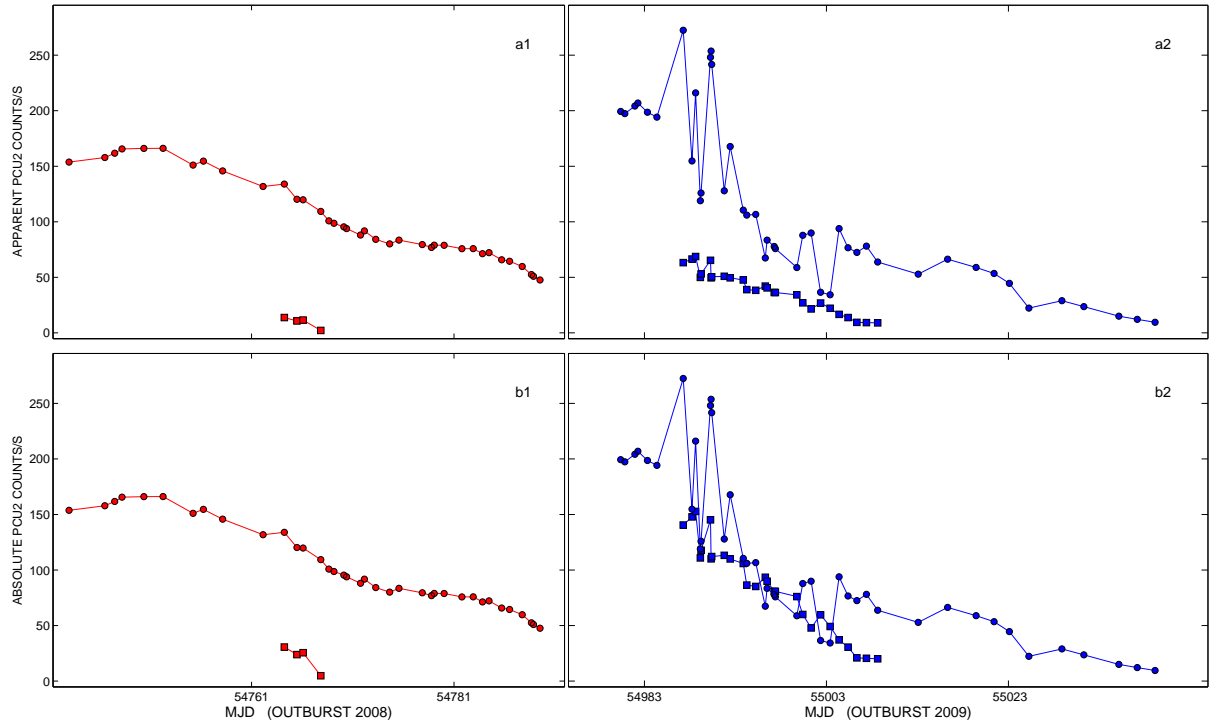}
\caption{Evolution of the count rates associated to the disk blackbody and to the powerlaw components counts/s as a function of the time for the 2008 (left panels) and 2009 (right panels). The circles track the powerlaw components, the squares track the disk-blackbody component . The upper panels show the lightcurves not corrected for the inclination angle effect. The lower panels show the lightcurves corrected for the inclination angle effect. For the 2009 outburst it is clear that, correcting for the inclination angle, the disk component becomes dominant in the soft state. }\label{fig:components}
\end{figure*}
%----------------------------------------------------------------------------------

\subsection{The flux evolution of the spectral components}
Two clear components are present in the BHT spectra. The disk-blackbody component (soft) usually dominates in the HSS (left part of the HID), while the powerlaw component dominates in the LHS (right part of the HID). During the intermediates states (HIMS and SIMS), both components are important. 

We calculated the fluxes (in the 2 - 200 keV energy band) related to the disk blackbody components and to the powerlaw components as a function of the time. For the latter we made a correction on the values by excluding the reflection contribution, which depends on the inclination angle of the source $\Theta$. The count rates associated to both the powerlaw and the disk-blackbody component are plotted in Fig. \ref{fig:components} (upper panels). It is clear that the count rate is always powerlaw dominated. This is consistent with the fact that during the 2008 outburst the source basically never shows a strong disk component. However, during the 2009 outburst the source clearly reached the HSS and yet the disk never became dominant. We interprete the lack of soft emission in terms of the inclination of the source. 
%H1743-322 is thought to be at high inclination angle (see \citealt{MacC2009}). 

Assuming a spherical geometry for the corona, the amount of the hard photons we received can be roughly considered inclination independent. The disk emission is expected to be non-isotropic and should be corrected taking into account the inclination of the source. 
We calculated the minimum inclination angle to obtain a disk count rate higher than the powerlaw count rate in the soft state. We performed this operation on the softest observation (Obs. \#27 of the 2009 outburst), assuming that the observed disk emission is $cos(\Theta$) fainter than the real one. We obtained a minimum angle of $\sim$ 43$^o$, which is consistent with the values reported by \cite{MacC2009} ($\sim$70$^{\circ}$). 

Applying the inclination angle correction to all the 2009 observations (assuming $\Theta $= 65$^o$ we obtain the values shown in Fig. \ref{fig:components}, panel b2. During the 2009 outburst, the disk probably becomes dominant in the central part of the outburst, in correspondence of the HSS. Given that the disk dominance in this state is usually more evident, it is possible that the inclination angle is even larger than 65$^o$. Since the system does not show eclipses and the inclination is in the range $\sim$43$^o$ - 75$^o$, probably the lower limit is unrealistically low because, assuming an inclination of $\sim$ 43$^o$, only the softest observation would be dominated by the disk.
No big differences are observed when correcting 2008 data from inclination effects (see Fig. \ref{fig:components}). This reinforces the statement that during the 2008 outburst the disk emission never dominates the spectrum.

%%%%%%
\section{Conclusions}
The different behavior shown by H1743 during these two events has allowed us to study the evolution of the spectral parameters at different accretion regimes.
During the 2009 outburst the system followed the canonical evolution through all the states usually seen in BHTs. In the 2008 outburst only the hard states are clearly reached and we could state, from the timing analysis, that the source did not reached the soft states while observed by RXTE. We find that the energy spectrum of the 81 observations we have analyzed can be described by using a model consisting of a disk black-body, a powerlaw + reflection component, an absorption and a gaussian line component. The spectral parameters derived by using this model are acceptable from a physical point of view and consistent between each other. The evolution of the spectral parameters is consistent between the two outbursts, and it does not allow to predict the subsequent behavior of the source. 

We conclude that this different behavior cannot be predicted on the basis of the initial spectral parameter evolution. The occourring of the transion is possibly driven by a parameter independent by the spectral properties of the source, even thought a scenario in which the transition took place while the source was not observed cannot be ruled out.

\vspace{1cm}
\noindent SM and TB acknowledge support to the ASI grant I/088/06/0. 
The research leading to these results has received funding from the European Community's Seventh Framework Programme (FP7/2007-2013) under grant agreement number ITN 215212 \textquotedblleft Black Hole Universe\textquotedblright.  
\bibliographystyle{mn2e.bst}
\bibliography{bib_H1743.bib} 
\label{lastpage}
\end{document}